\documentclass[twoside,11pt]{article}

\usepackage[preprint]{jmlr2e}

\usepackage{latexsym}
\usepackage{amsmath,amssymb,amsfonts,bm,amsbsy,bbm}
\usepackage{graphicx,rotating}
\usepackage{multirow,color}
\usepackage{physics}
\usepackage{tabularx,booktabs}
\newcolumntype{Y}{>{\scriptsize\raggedright\arraybackslash}X}
\usepackage{mathtools}
\usepackage{makecell}
\usepackage{subcaption}
\usepackage{listings}
\lstdefinestyle{Rstyle}{
  language=R,
  basicstyle=\small\ttfamily,
  keywordstyle=\color{blue},
  commentstyle=\color{green!60!black},
  stringstyle=\color{red},
  numbers=left,
  numberstyle=\tiny\color{gray},
  stepnumber=1,
  numbersep=5pt,
  backgroundcolor=\color{gray!10},
  frame=single,
  showspaces=false,
  showstringspaces=false,
  showtabs=false,
  tabsize=2,
  breaklines=true,
  breakatwhitespace=false,
  captionpos=b
}

\newcommand{\E}{E}

\def\bse{\begin{eqnarray*}}
\def\ese{\end{eqnarray*}}
\def\be{\begin{eqnarray}}
\def\ee{\end{eqnarray}}
\def\bsq{\begin{equation*}}
\def\esq{\end{equation*}}
\def\bq{\begin{equation}}
\def\eq{\end{equation}}
\def\bi{\begin{itemize}}
\def\ei{\end{itemize}}

\def\hat{\widehat}

\def\shared{{\boldsymbol\psi}}

\def\efree{\boldsymbol\varphi}
\def\bGamma{{\boldsymbol\Gamma}}

\def\bdelta{{\boldsymbol\delta}}
\def\bSigma{{\boldsymbol\Sigma}}

\def\ba{{\boldsymbol\zeta}}
\def\bphi{{\boldsymbol\phi}}

\def\pfree{{\boldsymbol \vartheta}}

\def\bgamma{{\boldsymbol\gamma}}

\def\btheta{\boldsymbol\theta}
\def\0{{\bf 0}}
\def\D{{\boldsymbol D}}
\def\X{{\boldsymbol X}}
\def\x{{\boldsymbol x}}

\def\Z{{\boldsymbol Z}}
\def\z{{\boldsymbol z}}

\def\U{{\boldsymbol U}}
\def\u{{\boldsymbol u}}

\let\vaccent\v
\def\v{{\boldsymbol v}}
\def\S{{\boldsymbol S}}

\def\B{{\boldsymbol B}}
\def\b{{\boldsymbol b}}
\def\A{{\boldsymbol A}}
\def\a{{\boldsymbol a}}

\def\q{{\boldsymbol q}}

\def\G{{\boldsymbol G}}
\def\H{{\boldsymbol H}}

\def\I{{\boldsymbol I}}

\def\C{{\boldsymbol C}}
\def\J{{\boldsymbol J}}
\def\L{{\boldsymbol L}}
\def\R{{\boldsymbol R}}

\def\t{{\boldsymbol \lambda}}
\def\0{{\bf 0}}
\def\1{\mathbb{1}}
\def\mle{\hat\btheta_{{}_{\rm MLE}}}
\def\fmle{\hat\btheta_{{}_{\rm FMLE}}}

\def\var{\mathrm{Var}}

\usepackage{boxedminipage}

\newtheorem{assumption}{Assumption}

\usepackage{tikz}
\usetikzlibrary{shapes, arrows}

\usepackage{lastpage}

\jmlrheading{1}{2025}{1-\pageref{LastPage}}{}{}{}{Chi-Shian Dai and Jun Shao}

\ShortHeadings{Fused Multinomial Logistic Regression}{Dai and Shao}
\firstpageno{1}

\begin{document}

\title{Fused Multinomial Logistic Regression Utilizing  Summary-Level External Machine-learning Information}
\author{\name Chi-Shian Dai \email cdai@gs.ncku.edu.tw\\
\addr Department of Statistics\\
 National Cheng Kung University\\
 Tainan, 701, Taiwan
\AND
\name Jun Shao\,$^*$ \email shao@stat.wisc.edu\\
\addr Department of Statistics\\
University of Wisconsin-Madison\\
 Madison, 53706, USA}

\editor{}

\maketitle

\renewcommand{\thefootnote}{\fnsymbol{footnote}}
\footnotetext[1]{Corresponding author}
\renewcommand{\thefootnote}{\arabic{footnote}}

\begin{abstract}%
In many modern applications, a carefully designed primary study provides individual-level data for interpretable modeling, while summary-level external information is available through black-box, efficient, and nonparametric machine-learning predictions. Although summary-level external information has been studied in the data integration literature, there is limited methodology for leveraging external nonparametric machine-learning predictions to improve statistical inference in the primary study.
We propose a general empirical-likelihood framework that incorporates external predictions through moment constraints. An advantage of nonparametric machine-learning prediction is that it induces a rich class of valid moment restrictions that remain robust to covariate shift under a mild overlap condition without requiring explicit density-ratio modeling. 
We focus on multinomial logistic regression as the primary model and address common data-quality issues in external sources, including coarsened outcomes, partially observed covariates, covariate shift, and heterogeneity in generating mechanisms known as concept shift. We establish large-sample properties of the resulting fused estimator, including consistency and asymptotic normality under regularity conditions. Moreover, we provide mild sufficient conditions under which incorporating external predictions delivers a strict efficiency gain relative to the primary-only estimator. Simulation studies and an application to the National Health
and Nutrition Examination Survey on multiclass blood-pressure classification. Code is available at \url{https://github.com/chichiihc2/MLfused.git}.\end{abstract}


\section{Introduction}

In recent years, researchers have increasingly moved beyond analyzing a single dataset toward integrating multiple data sources to improve statistical efficiency. In many applications, a primary study is carefully designed and provides high-quality individual-level data with a not-so-large sample size, while additional external information is available from auxiliary sources with much larger sample sizes but
only summary statistics (not individual-level data). A growing body of literature has investigated how to incorporate summary-level external information 
\citep{chatterjee2016constrained, huang2016efficient,zhang2017statistical, sheng2020censored,zhang2020generalized,zheng2022risk,cheng2023semiparametric,ding2023fitting,gao2023noniterative,gu2023synthetic,chiKernel2024,shao2024gmm,fang2025integrated}.

In this paper, we focus on multiclass classification as the primary inferential task. We adopt multinomial logistic regression, which, in addition to prediction, provides a principled inferential framework in which regression coefficients are interpretable as log-odds ratios or contrasts for comparison. To enhance multinomial logistic regression without sacrificing its interpretability, we use extra moment conditions constructed from external summary-level predictors from modern machine-learning methods \citep{hastie2009elements} such as gradient boosting, XGBoost, regression trees,  random forests, and deep neural networks, which have achieved remarkable predictive success with large training datasets. These machine-learning methods are  powerful and robust (nonparametric), which induces a rich class of valid moment conditions, unlike 
parametric methods in external sources that are often not robust against model violations. 
Our work mainly bridges two complementary sources of information: an interpretable multinomial logistic regression and a powerful, robust, but black-box external predictor to improve efficiency. 

The main challenge in leveraging external information is the heterogeneity across data sources, including covariate shift (the differences in covariate populations between the primary and external sources) and heterogeneity in outcome-generating mechanisms (the differences in conditional means of outcomes given covariates in different sources), also known as concept shift or drift \citep{moreno2012unifying,gama2014survey}.  

Covariate shift can be handled by estimating density ratios 
\citep{chiKernel2024} when external individual-level data are available, or when only summary-level external information is available through some models on density ratios \citep{cheng2023semiparametric,gu2023synthetic,shao2024gmm} or through moment selection  \citep{fang2025integrated}. In our approach, we handle covariate shift without estimating the density ratio, assuming that unmeasured covariates (if any) in the external source are missing at random and leveraging the fact that machine-learning methods are non-parametric, which are robust to covariate shift.

The heterogeneity in outcome-generating mechanisms can arise from differences in enrollment,
sampling, or case mix, which in turn can alter class proportions. To accommodate this, we divide the regression parameters into two sets: a set of
free parameters representing source-specific discrepancies and another set of shared parameters allowing information transfer across different data sources. 
Without shared parameters, the primary and external sources are disconnected, and external information does not help. To ensure the success of transferring external information, the number of free parameters cannot exceed the effective number of moment conditions contributed by the external information. We discuss how to construct enough moment conditions after developing the methodology and related asymptotic theory.

Our main contributions are fourfold. First, we propose a general data fusion/integration framework that incorporates external nonparametric machine-learning probability predictions into a primary, interpretable
multiclass classification model. Second, we develop ideas to address
heterogeneity in covariates and outcome-generating mechanisms
across different sources. Third, the proposed framework can handle the external sources with partial covariates and coarsened labels. Lastly, we establish large-sample properties of the resulting fused estimator, including consistency and asymptotic normality under regularity conditions, and we derive mild conditions under which incorporating external predictions yields a strict efficiency gain relative to the primary-only estimator.

 The paper is organized as follows. Section~\ref{sec:data.structure} introduces the notation and formalizes the data structure arising from heterogeneous primary and external sources.
Section~\ref{sec:methodology} develops the proposed fused estimation methodology and establishes its key theoretical properties.
Section~\ref{sec:simulation} evaluates the finite-sample performance of the
proposed estimator through simulation studies, demonstrating clear efficiency gains over methods that do not use external information.
Section~\ref{sec:Real_Data} illustrates the practical utility of the proposed approach using a real data example from the National Health and Nutrition Examination Survey, focusing on a multiclass blood pressure classification problem. Section~\ref{sec:Discussion} provides a discussion. The Appendix contains all technical proofs.

\section{Data Structure}\label{sec:data.structure}

We introduce the data structures for a primary study ($S=1$) and one external study ($S=0$). Extensions to multiple external studies are straightforward. 

\subsection{Primary Study}\label{sec:prim}
Let $(Y_i, \X_i)$, $i = 1, \ldots, n$, denote $n$ independent and identically distributed observations from $(Y, \X )$ under the primary study ($S=1$),  where $Y \in \{1, \ldots, K\}$ is a class label outcome,  
$\X $ is a $p$-dimensional covariate vector  whose first component is 1 (corresponding to an intercept) and the remaining  components are observed features, 
and $K \geq 2$ and  $p \geq 2$ are fixed and known.
For the outcome-generating mechanism, we assume that, 
conditional on $\X$,  the class label $Y$ follows the multinomial logistic regression model,
\begin{equation}\label{outcome-primary}
 P(Y=k \mid \X, S=1)
= \left\{ \begin{array}{ll}
\frac{\exp(\X^\top \btheta_k)}
{1+ \sum_{j=1}^{K-1} \exp(\X^\top \btheta_j)} & \qquad
 k=1,\ldots,K-1, \vspace{2mm} \\[2mm]
\frac{1}
{1+ \sum_{j=1}^{K-1} \exp(\X^\top \btheta_j)} & \qquad k=K , \end{array} \right.
\end{equation}
where $\btheta_1,...,\btheta_{K-1}$ are source-specific regression parameters and, throughout, $\x^\top$ denotes the transpose of vector $\x$. The target parameter to be estimated is
$\btheta = (\btheta_1^\top , ..., \btheta_{K-1}^\top )^\top$.

\subsection{External Source}\label{sec:observed}

Let $(U_i, \Z_i)$, $i = 1, \ldots, N$, denote $N$ independent and identically distributed observations from $(U,\Z )$ under an external study ($S=0$).
The external sample size $N$ is much larger than the sample size $n$ of the primary study in the sense that $n/N \to 0$ as $n$ grows to $\infty$.
In the ideal setting, $(U,\Z)$ has the same form as $(Y,\X)$ in the primary study, although their populations may be different. 
In many applications, however, the external study may
record only a subset of covariates, $\Z\subseteq \X$ (for example, due to 
availability or privacy restrictions) and/or coarsened outcome label
(for example, collapsed or partially observed categories)
$U \in \{1,..., L\}$ with $U = l$ if $Y \in {\cal C}_l $, where ${\cal C}_l$ is a subset of $\{1,..., K\}$, ${\cal C}_l \cap {\cal C}_{l'} = \emptyset$ for $l \neq l'$, and ${\cal C}_1\cup \cdots \cup {\cal C}_L \subseteq \{1,..., K\}$ (some classes may be absent from the external study). 

The individual values of $(U_i,  \Z_i) $ in the external source are not available to help the analysis of primary data. What is available from the external source is a nonparametric machine-learning  prediction denoted by 
$$\hat \q (\z ) = \Bigg(\sum_{k\in C_1} \hat q_k(\z),\ldots , \sum_{k\in C_{L-1}}\hat q_k(\z)\Bigg)^\top,$$ 
which gives an estimator of the true population probability vector 
$$\q (\z)= \big( P(U=1  \mid \z , S=0 ), \ldots , P( U=L-1 \mid \z, S=0 )  \big)^\top .$$ 
A prediction of the outcome label associated with $\z$ can be obtained from $\hat \q (\z)$ for any $\z$. 
 
The external machine-learning predictor $\hat \q$ is constructed using $(U_i,\Z_i)$, $i=1,..., N$, as training data (although they are not available for primary data analysis). Examples of external machine-learning procedures include the nearest neighbors regression, 
regression trees, kernel regression, and 
more advanced methods such as generalized random forests, gradient boosting, XGBoost, and deep neural networks.
The prediction rule $\hat \q$ is available in a black-box manner to compute $\hat \q (\z)$; in fact, we do not even need to know what exact procedure was used to construct $\hat \q$.

For each $\z$, let $\overline\q (\z)  - \q (\z )$ be the bias of $\hat \q (\z )$, where $\overline\q (\z) = E \{ \hat \q (\z) \mid \z \}$.  
 The following assumption is for the asymptotic validity of $\hat \q$. 

\begin{assumption} As $n \to \infty$ and $N \to \infty$, \vspace{1mm}

\noindent
(i)  $\|\hat{\q}-\q \|_\infty \xrightarrow{p} 0$, where 
$\| \cdot \|_\infty$ is the sup-norm and $\, \xrightarrow{p} \, $  is convergence in  probability;\vspace{1mm}

\noindent
(ii)   $\sqrt{n} \, \|\overline{\q}-\q \|_\infty \xrightarrow{p} 0$ or $n \, E \| \overline{\q}(\Z) - \q (\Z) \|^2  \rightarrow 0$, where $\| \cdot \|$ is the Euclidean norm;\vspace{1mm}

\noindent
(iii) $ E \{ \sigma_l^2 ( \Z )h^2(\Z)\} \to  0$  for any $l$ and integrable $h^2$, where 
$   \sigma_l^2(\z) \!= \! \var \{ \sum_{k\in C_l} \hat  q_k(\z)\mid \z \}$. 
 \end{assumption}

Assumption 1(i) is the uniform consistency of $\hat \q$ as an estimator of $\q$ and is  typically true for nonparametric machine-learning methods, since under standard conditions \citep{stone1982optimal},  $
  \|\hat{\q}-\q \|_\infty   =
  O_p\!\big( (\log N/N)^{1/(2+p/\alpha)}\big)$, 
where $O_p(a_N)$ is a term bounded by $a_N$ in probability and 
$\alpha>0$ measures the smoothness of $\q$. 
Since $\overline\q - \q$ is the bias of $\hat \q$, Assumption 1(ii) simply says that $\hat \q$ is asymptotically valid in terms of bias. Typically $N^\tau  \|\overline{\q}-\q \|_\infty \xrightarrow{p} 0$ for some $\tau \leq 1/2$. If $N$ is of the order  $n^\eta$ for some $\eta > 1 $, then Assumption 1(ii) holds when $\eta \tau \geq 1/2$. 
The same discussion applies when $ \|\overline{\q}-\q \|_\infty $ is replaced by $\{E \| \overline{\q}(\Z) - \q (\Z) \|^2 \}^{1/2}$. 

Assumption 1(iii) means that the variability of the external predictor is under control, which holds 
for a broad class of nonparametric learners. 
For example, for $\kappa$-nearest neighbors regression, typically 
$\sigma_l^2(\Z) \propto \kappa^{-1}$,
reflecting the averaging over $\kappa$ local neighbors.
For regression trees, the prediction at $\Z$ is an average of observations falling in the same terminal node (leaf), yielding
$
\sigma_l^2(\Z) \propto (\text{the number of observations in the leaf having }\Z)^{-1}$.
For $d$-dimensional kernel regression with bandwidth $b$ and external sample size $N$, the standard variance calculation gives
$\sigma_l^2(\Z) \propto (N b^{d})^{-1}$,
where $Nb^d$ is the effective number of observations within the kernel window.
For more advanced ensemble methods such as generalized random forests, the prediction variance $\sigma_l^2 (\Z )$ is approximately of order
$ s/N$,
where $s$ is the subsample size used to build each tree~\citep{athey2019generalized}.
Together, these examples suggest that Assumption 1(iii) is mild and practically plausible, as $n/N \to 0$.

\subsection{Heterogeneity and Connection between Primary and External Studies} \label{sec:connection}

Heterogeneity between the primary and external data populations typically exists. Covariate shift refers to the difference between the population distribution of $\X$ from the primary study and that of $\Z$ from the external source. Unlike in previous studies, we do not impose any assumption on covariate shift when $\Z = \X$. 
When $\Z \neq \X$, we assume that components in $\X$ but not in $\Z$ are omitted at random; see Assumption 3 in Section  \ref{sec:moment}, where we also explain why we need this assumption. 

For the outcome-generating mechanism of the external source, we assume the same type of multinomial logistic regression model when $(U,\Z)$ has the same form as $(Y,\X)$,  although in applications, $U$ may be coarsened and $\Z$ may be a subset of $\X$: 
\begin{equation}\label{outcome-external}
 P(Y=k \mid \X, S=0)
= \left\{ \begin{array}{ll}
\frac{\exp(\X^\top \bphi_k)}
{1+ \sum_{j=1}^{K-1} \exp(\X^\top \bphi_j)} & \qquad
 k=1,\ldots,K-1, \\[2mm]
\frac{1}
{1+ \sum_{j=1}^{K-1} \exp(\X^\top \bphi_j)} & \qquad k=K , \end{array} \right.
\end{equation}
where $\bphi_k$'s are unknown parameters and 
$\bphi = (\bphi_1^\top , ..., \bphi_{K-1}^\top )^\top$ can be different from  $\btheta = (\btheta_1^\top , ..., \btheta_{K-1}^\top )^\top$  in the primary study given in (\ref{outcome-primary}). Note that
 $\sum_{k \in {\cal C}_l} P(Y=k \mid \z, S = 0)=P( U=l \mid \z, S=0) $  in Section \ref{sec:observed}. 

Although we allow heterogeneity between outcome-generating mechanisms (\ref{outcome-primary}) and (\ref{outcome-external}), that is, $\btheta$ and $\bphi$ are distinct, 
if they are totally unrelated, then
the two sources are disconnected, and external information cannot
be used to improve the estimation of the primary target $\btheta$. Thus, to borrow strength
from the external source, 
we impose the following structural assumption for the connection between the two sources.
\begin{assumption}\label{connection}
The parameter $\btheta$ in \eqref{outcome-primary} and $\bphi$ in \eqref{outcome-external} satisfy 
$\btheta = \big( \shared^\top, \ \pfree^\top  \big)^\top  $ and  
$\bphi = \big( (\A_m \shared )^\top , \ \efree^\top \big)^\top $,
where $\A_m$ is a known  $m \times m$ matrix, 
$\shared$ is an $m$-dimensional shared parameter vector transporting information available from the external source,  $\pfree$, and 
$\efree$ are $\{p(K-1)-m\}$-dimensional primary and external free parameters, respectively,  $0 \le m \leq p(K-1)$, and $\A_m$ is invertible when $m>0$.  
\end{assumption}

The following are two examples. \vspace{2mm}

\noindent
{\bf Example 1} (Full transportability).
If $\btheta=\bphi$, then the two sources are fully aligned. In this case, the shared component is the entire parameter vector, that is,
$\shared=\btheta = \bphi$, and Assumption \ref{connection} holds with $\A_m =$  the
identity matrix of dimension $m=p(K-1)$. \vspace{2mm}

\noindent
{\bf Example 2} (Proportion heterogeneity). 
In many classification applications, marginal class proportions differ across data sources due to differences in enrollment, sampling, or case-mix. Under
multinomial logistic regression, such shifts are often well approximated by allowing intercept terms to differ while keeping slope coefficients invariant.
Specifically, if we write
\[
\btheta_k=(\theta_{k,1},\theta_{k,2},\ldots,\theta_{k,p})^\top ,
\qquad 
\bphi_k=(\phi_{k,1},\phi_{k,2},\ldots,\phi_{k,p})^\top,
\qquad k=1,\ldots,K-1,
\]
where the first coordinate corresponds to the intercept, then Assumption \ref{connection} holds with 
\begin{align*}
\shared
& =
(\theta_{1,2},\ldots,\theta_{1,p},\ \theta_{2,2},\ldots,\theta_{2,p},\ \ldots,\ \theta_{K-1,2},\ldots,\theta_{K-1,p})^\top, \\
\pfree & = (\theta_{1,1}, \theta_{2,1},\ldots,\theta_{K-1,1} )^\top ,\quad
\efree  = (\phi_{1,1}, \phi_{2,1},\ldots,\phi_{K-1,1} )^\top ,
\end{align*}
and  $\A_m = $ the identity matrix of dimension  $m=(p-1)(K-1)$.  In this
setting, covariate log-odds ratios are shared across sources, while the baseline
prevalence (captured by the intercepts) is allowed to differ.

\section{Methodology and Theory}\label{sec:methodology}

Estimation of the target parameter $\btheta$ in \eqref{outcome-primary} is essential for prediction of $Y$ or inference on $\btheta$. 
With primary data alone, the standard maximum likelihood estimator (MLE) $\mle$ of $\btheta$ is obtained by maximizing the log-likelihood
\begin{equation}
\ell_n(\btheta ) = \frac{1}{n} \sum_{i=1}^{n} \sum_{k=1}^{K-1} 1(Y_i=k) \log p_k(\X_i \mid \btheta ) \label{primary-likelihood} 
\end{equation}
over $\btheta$, where $p_k(\X \mid\btheta )$ is the right side of (\ref{outcome-primary}) and $1( \cdot )$ is the indicator function. 
Our approach is to apply the  empirical likelihood \citep{owen2001empirical,qin2000miscellanea} with external information 
used as constraints added to maximizing \eqref{primary-likelihood} to gain estimation efficiency. 

\subsection{Methodology}\label{sec:moment}

In the easy case where the external source does not have omitted covariates ($\Z= \X$) and coarsened outcome labels, with the notation in Section \ref{sec:observed}, 
$\q (\X) = \big( P(Y = 1 \mid \X, S=0), \ldots , P(Y=K-1 \mid \X, S=0 ) \big)^\top$
has the $k$th component $q_k (\X) = p_k(\X\mid \bphi)$, the right side of  \eqref{outcome-external}. 
Therefore,  to gain efficiency in estimating the target $\btheta$ in the primary study, we can add the following constraints based on the primary sample, 
\begin{equation}\label{eq:Empirical_moments}
\sum_{i=1}^n \delta_i \,
\{ p_k(\X_i \mid \bphi ) - \hat q_k(\X_i)\}
= 0, \qquad k=1,...,K-1, 
\end{equation}
where 
$\hat q_k(\X)$ is the external
machine-learning estimate  of $q_k(\X )= P( Y=k \mid \X , S=0)$ defined in Section \ref{sec:observed} 
when the outcome is not coarsened and $\Z = \X$,
and $\delta_i$'s are non-negative weights satisfying $\sum_{i=1}^n \delta_i =1$.

As we discussed in Section \ref{sec:observed}, the external source covariate is often a sub-vector $\Z \subset \X$, instead of the entire $\X$ in the primary study. 
In this case, in order to use constraints \eqref{eq:Empirical_moments} with $\hat q_k (\X_i )$ replaced by $\hat q_k (\Z_i)$, we need the moment condition
\begin{equation}\label{eq:adv}
    E \{ p_k( \X \mid \bphi )  - q_k ( \Z ) \mid S=1 \} =0, 
\end{equation}
where $q_k ( \Z )
 = P( Y=k \mid \Z , S=0) = E \{ P( Y=k \mid \X  , S=0) \mid \Z , S=0  \}$.   However, \eqref{eq:adv} may not hold  when 
the density ratio $f_1(\X)/f_0(\X)$ is a function of the entire vector $\X$ (see the Appendix), where $f_1$ and $f_0$ are the densities of $\X$ in the primary and external sources, respectively. An assumption on $f_1(\X)/f_0(\X)$ is required to connect the two sources, that is, to ensure \eqref{eq:adv}.
One such assumption is a ratio model \citep{cheng2023semiparametric,gu2023synthetic,shao2024gmm}, but  
$f_1(\X)/f_0(\X)$ under the ratio model needs to be estimated, which requires some additional information from the external source or individual-level external data, and is sensitive to the choice of ratio model. 

Instead, we make the following assumption and avoid the estimation of $f_1(\X)/f_0(\X)$. 

\begin{assumption}\label{asm:ignorability}
The ratio 
$  f_1(\X)/ f_0(\X) 
$
is a function of $\Z$. 
\end{assumption}

Assumption 3 is
closely related to the missing at random condition in the missing data
literature.  In other words,
if the component of $\X$ not in $\Z$
is considered a missing covariate, then Assumption~\ref{asm:ignorability} means that the missingness is at random, that is, the missing covariate and the indicator $S$ are independent conditioned on the observed $\Z$.
 
Under Assumption 3, \eqref{eq:adv} holds  (which is shown in the Appendix) and, hence, without estimating the density ratio $f_1(\X)/f_0(\X)$, we can still use 
\eqref{eq:Empirical_moments} with $\hat q_k(\X_i)$ replaced by $  \hat q _k(\Z_i )$ given in Section  \ref{sec:observed} when the outcome is not coarsened. 

Before we present the likelihood using constraints given by \eqref{eq:Empirical_moments}, we want to add the following two components. 

First, a square integrable function $h$ can be added to  
\eqref{eq:adv}, that is, 
\begin{equation}\label{eq:adv+}
    E\bigl[\{p_k(\X\mid \bphi) -q_k( \Z ) \} \, h( \Z) \mid S=1\bigr] 
= 0 .\tag{5+} 
\end{equation}  
Adding $h$ is mainly because of gaining efficiency, as we discuss later (in Theorem \ref{coro:gain} of Section \ref{sec:eff_fused} and afterward). 
If we consider parametric likelihood  \eqref{primary-likelihood}  under the primary study, then the derivative of $\ell_n ( \btheta )$ naturally leads to $h (\X ) = \X$. 
Since the external source provides nonparametric machine-learning $\hat \q$, we can have a more flexible choice of $h$. However, an optimal $h$, even if it exists, is not easy to construct since it likely depends on unknown quantities. Instead, we propose to consider a class ${\cal H}$ of finitely many $H$ 
 base functions and replace constraint \eqref{eq:Empirical_moments} 
by 
\begin{equation}
\sum_{i=1}^n \delta_i \,
 \{ p_k(\X \mid \bphi ) - \hat q_k(\Z)\} h(\Z)
= 0, \qquad k=1,...,K-1,\qquad h\in \mathcal{H} ,  \label{eq:Empirical_moments1}
\end{equation} 
with $(K-1) H >$ the dimension of free external parameter $\efree$ in Assumption 2 to gain efficiency, according to our discussion after Theorem \ref{coro:gain}.  
Specifically, we may let  $\mathcal{H}$ contain all components of $\Z$; if we need more functions, we may consider 
a natural cubic spline
basis (for each component of $\Z$) with a small number of interior knots placed at empirical quantiles. 

Second, in many applications, the primary study records a fine-grained outcome label
$Y\in\{1,\ldots,K\}$, but the external source provides only a coarsened label
$U=l$ if $ Y\in {\cal C}_l$, $l=1,\ldots,L$,
  ${\cal C}_l\cap {\cal C}_{l'}=\emptyset$ for $l\neq l'$, and
${\cal C}_1 \cup \cdots \cup {\cal C}_L \subseteq \{1,\ldots,K\}$, and 
we can only observe the grouped machine-learning prediction $\sum_{k\in {\cal C}_l}\hat q_k(\Z)$, instead of $\hat q_k(\Z)$ for each $k$. 
In this scenario, 
we can use constraint \eqref{eq:Empirical_moments1} 
with $p_k(\X\mid\bphi)$ replaced by $\sum_{k\in C_l} p_k(\X\mid\bphi)$ 
and $\hat q_k(\Z)$ replaced by $\sum_{k \in {\cal C}_l}\hat q_k(\Z)$.

Now we are ready to present the empirical likelihood to combine the primary multinomial likelihood with the external moment information constraints, that is, we estimate 
$\btheta = (\btheta_1^\top , ..., \btheta_{K-1}^\top )^\top$  in the primary study given in (\ref{outcome-primary}) by maximizing 
the Lagrangian log-pseudo-likelihood 
\begin{equation}
\ell_n(\btheta )
+ \frac{1}{n} \sum_{i=1}^{n} \log \delta_i
- \lambda_0 \left( \sum_{i=1}^{n} \delta_i - 1 \right)
- \sum_{h \in {\cal H}} \sum_{l=1}^{L-1} \sum_{i=1}^{n}
  \delta_i \,  g_{l,h} (\X_i \mid \A_m \shared , \efree  ) \, \lambda_{l,h} \label{likelihood}
\end{equation}
over 
$\btheta$, the external free parameter $\efree$ defined in Assumption \ref{connection}, $\delta_i$'s, and Lagrange multipliers
$\lambda_0$ and $\lambda_{l,h}$'s, 
where 
$\ell_n(\btheta ) $
is  log-likelihood \eqref{primary-likelihood}  using primary data only, $g_{l,h} (\X \mid \A_m \shared , \efree  )= \sum_{k \in {\cal C}_l} \{ p_k(\X \mid \bphi )  - \hat q _k(\Z )\}h (\Z )$, and $\shared$ and $\efree$ are given in Assumption \ref{connection}. 

Maximizing  (\ref{likelihood}) with respect to $\delta_i$'s  and  $\lambda_0$ yields 
 $\hat \delta_i=n^{-1} \{1 +
\sum_{h \in {\cal H}} \sum_{l=1}^{L-1} g_{l,h} (\X_i \mid \A_m \shared , \efree ) \, \lambda_{l,h} \}^{-1}$ and $\hat \lambda_0=1$, which leads to the following profile log-pseudo-likelihood:
\begin{align}
 \ell_n(\bgamma  \mid \hat \q \, ) = \ell_n ( \btheta ) - \frac{1}{n}
 \sum_{i=1}^n\log \left\{ 1 +
  \sum_{h \in {\cal H}} \sum_{l=1}^{L-1}  g_{l,h}(\X_i \mid \A_m \shared , \efree ) \, \lambda_{l,h}
  \right\}
\label{e1},
\end{align}
where $\bgamma
= \bigl( \t^\top , \btheta^\top, \efree^\top \bigr)^\top $
is the enlarged parameter vector with 
$\t = ( \lambda_{l,h}, l=1,...,L-1, h \in {\cal H} )^\top $.
 The fused maximum likelihood estimator $\hat{\bgamma}$ of $\bgamma$ is  given by
\begin{equation}\label{hat.eata}
   \hat{\bgamma}
= \arg\max_{\bgamma} \, \ell_n(\bgamma \mid \hat \q \, ).
\end{equation}
The fused maximum likelihood estimator (FMLE)  $\fmle$ of target parameter $\btheta$ in \eqref{outcome-primary} is then the sub-vector of $\hat{\bgamma}$ in \eqref{hat.eata} 
 corresponding to the estimation of $\btheta$.

\subsection{Consistency and Asymptotic Normality of  Fused Estimator}

We consider asymptotics as the primary study sample size $n \to \infty$ and $n/N \to 0$. Throughout, $\xrightarrow{p}$ and 
$\xrightarrow{d}$ denote respectively convergence in probability and convergence in distribution. 
The proofs of all theorems are given in the Appendix. 

Our first result is the consistency of fused
estimator $\hat{\bgamma}$ in \eqref{hat.eata}. 

\begin{theorem}[Consistency]\label{thm:consistency}
Under Assumptions 1(i), 2-3 and the regularity conditions (C1)-(C3) stated in the Appendix,
 any maximizer $\hat{\bgamma}$ of $\ell_n(\bgamma \mid \hat\q \, )$ is
consistent, 
that is, 
$ \hat{\bgamma} \;\xrightarrow{p}\; \bgamma_0 $, where $\bgamma_0$ is the true maximizer  of $E \{  \ell ( \bgamma \mid \q \, )\}$  given in condition (C1) and 
 $\ell (\bgamma \mid \q \, )$ 
 denotes  likelihood  \eqref{e1} with $n=1$ and $\hat \q$ replaced by $\q$.
\end{theorem}

We next turn to the asymptotic distribution of  $\hat{\bgamma}$. Since
$\hat{\bgamma}$ is an $M$–estimator, a standard argument
\citep{newey1994large} yields that
\begin{equation}\label{asy}
      \sqrt{n}(\hat{\bgamma} - \bgamma_0 )
  \, - \,
  \sqrt{n} \bigl\{\nabla^2_{\bgamma\bgamma}  \ell_n(\bgamma \mid \hat \q \,)\bigr\}^{-1}
 \,\nabla_{\bgamma} \ell_n(\bgamma \mid \hat \q \, ) \Big|_{\bgamma = \bgamma_0} \, \xrightarrow{p}\; 0,
\end{equation}
so the limiting distribution is governed by the behavior of the empirical
Hessian and the score evaluated at $\bgamma = \bgamma_0$, where $\nabla_{\a}$ denotes the gradient and  $\nabla^2_{\a\b} = \nabla_{\a} \nabla_{\b}$
 for any vectors $\a$ and $\b$. 

\begin{theorem}[Asymptotic normality]\label{thm:clt.regu}
Under Assumptions 1-3 and
the regularity conditions (C1)-(C6) in the Appendix,   
the fused estimator $\hat{\bgamma}$ in \eqref{hat.eata} is asymptotically
normal:
\[
  \sqrt{n}\bigl(\hat{\bgamma}-\bgamma_0 \bigr)
\;  \xrightarrow{d} \; 
  N\!\left(\mathbf{0},\,\bSigma_{\bgamma_0} \right),
\]
where 
$ \mathbf{0}$ denotes a zero matrix of appropriate dimension,
 $\bSigma_{\bgamma_0} = \G^{-1}_{\bgamma_0} \J_{\bgamma_0}\G^{-1}_{\bgamma_0}$, 
 \begin{equation}\label{avar}
\J_{\bgamma_0}  
=  \begin{pmatrix}
\J_{\t_0} & \mathbf{0} & \mathbf{0} \\
 \mathbf{0} & \J_{\btheta_0} &  \mathbf{0} \\
  \mathbf{0} &  \mathbf{0} & \mathbf{0} \\
\end{pmatrix} , \qquad \G_{\bgamma_0} = \begin{pmatrix}
  \G_{\t_0\t_0} &  \G_{\t_0\btheta_0} &\G_{\t_0 \efree_0} \\
  \G_{\btheta_0\t_0} & \G_{\btheta_0\btheta_0} & \mathbf{0} \\
  \G_{\efree_0 \t_0} &  \mathbf{0} & \mathbf{0}  \\
\end{pmatrix} ,      
 \end{equation}
$\J_{\t_0} \!= \! \var \{ \nabla_{\t} \ell (\bgamma \mid \q \, ) \big|_{\bgamma = \bgamma_0}\}  $, 
$\J_{\btheta_0} \!= \! \var \{ \nabla_{\btheta} \ell (\bgamma \mid \q \, ) \big|_{\bgamma = \bgamma_0}\} $,
 $\G_{\a \b} = - E \, \{ \nabla^2_{\a \b} \ell (\bgamma \mid \q \, )\big|_{\bgamma=\bgamma_0}\}$ 
with $\a$ and $\b$ being appropriate components of $\bgamma$, 
$\G_{\a\b} = \G_{\b\a}^\top$, $\ell (\bgamma \mid \q \, )$ denotes likelihood  \eqref{e1} with $n=1$ and $\hat \q$ replaced by $\q$, and $\bgamma_0
= \bigl( \t_0^\top , \btheta_0^\top, \efree_0^\top \bigr)^\top $. Furthermore, $\J_{\t_0} = - \, \G_{\t_0\t_0} $ and $\J_{\btheta_0} = \G_{\btheta_0\btheta_0}$. 
\end{theorem}

\subsection{Asymptotic Efficiency of FMLE of Target Parameter}\label{sec:eff_fused}

For the estimation of the target parameter $\btheta$ in \eqref{outcome-primary},
we now consider the asymptotic relative efficiency between the fused estimator, FMLE $\fmle$ (the sub-vector of $\hat\bgamma $ in \eqref{hat.eata} corresponding to the estimation of $\btheta$), and the standard MLE
$\mle$ that maximizes likelihood \eqref{primary-likelihood} with
 data from the primary study alone. Let $\btheta_0$ be the sub-vector in $\bgamma_0
= \bigl( \t_0^\top , \btheta_0^\top, \efree_0^\top \bigr)^\top $.
A standard result is
\[ \sqrt{n} ( \mle - \btheta_0 ) \; \xrightarrow{d} \; N(\mathbf{0}, \I_{\btheta_0}^{-1}) , \]
where $\I_{\btheta_0} = \J_{\btheta_0}$ is the Fisher
information matrix in the primary study. 
Let $\bSigma_{\btheta_0}$ be the sub-matrix of 
$\bSigma_{\bgamma_0}$ in Theorem \ref{thm:clt.regu} corresponding to the asymptotic covariance matrix of FMLE $\fmle$ considered as a sub-vector of $\hat\bgamma $. 
Our discussion focuses on when 
$\fmle$ is asymptotically at least as efficient as $\mle $ in the sense that
\begin{equation}\label{eq:eff}
\bSigma_{\btheta_0} \preceq \I_{\btheta_0}^{-1}  ,
\end{equation}
where  $\A\preceq \B$ means that $\B-\A$ is positive
semi-definite for matrices $\A$ and $\B$. 

Our next theorem gives a more detailed form of 
$\bSigma_{\btheta_0}$. It also shows that \eqref{eq:eff} is actually achieved under the conditions in Theorem \ref{thm:clt.regu}. 

\begin{theorem}\label{thm:eff1}
Under the conditions of Theorem \ref{thm:clt.regu},
\begin{equation}\label{Sigma}
    \bSigma_{\btheta_0}  =  \I_{\btheta_0}^{-1}
     + \L_{\t_0\btheta_0}^\top \D_{\t_0} \L_{\t_0\btheta_0} , 
\end{equation} 
where  $
 \D_{\t_0} = - \, \J_{\t_0} - \G_{\t_0\btheta_0} \I_{\btheta_0}^{-1} \G_{\btheta_0 \t_0}$, 
$$    \L_{\t_0\btheta_0}  = \{\D_{\t_0}^{-1}-\D_{\t_0}^{-1}\G_{\t_0\efree_0}\bigl(\G_{\efree_0\t_0}\D_{\t_0}^{-1}\G_{\t_0\efree_0}\bigr)^{-1}\G_{\efree_0\t_0}\D_{\t_0}^{-1} \} \G_{\t_0\btheta_0}\I_{\btheta_0}^{-1}, $$
and  $\G_{\a \b}$ is as given in \eqref{avar}. As a result, \eqref{Sigma} implies \eqref{eq:eff}, since $ \D_{\t_0} $  is negative definite. 
\end{theorem}

If \eqref{eq:eff} holds but $\bSigma_{\btheta_0} \neq \I_{\btheta_0}^{-1}$, then incorporating external machine-learning predictions yields some asymptotically more efficient linear combinations of fused estimator $\fmle$ than the same linear combinations of $\mle$ based on primary data alone. If $\bSigma_{\btheta_0} = \I_{\btheta_0}^{-1}$, then the external information does not help in gaining efficiency. Because $\D_{\t_0}$ is negative definite, \eqref{Sigma} implies that $\bSigma_{\btheta_0} = \I_{\btheta_0}^{-1}$ if and only if $\L_{\t_0\btheta_0} = \0$. 

It is not simple to explain what $ \L_{\t_0\btheta_0}  = \0$ means, given the lengthy formula of the matrix  $ \L_{\t_0\btheta_0} $ in \eqref{Sigma}. 
The following result provides a necessary and sufficient condition for $ \L_{\t_0\btheta_0} = \0$ ($\bSigma_{\btheta_0} = \I_{\btheta_0}^{-1}$), which
 provides an insightful interpretation about when the external information provides no additional efficiency gain.  It also leads to discussions of some necessary and sufficient conditions for efficiency gain.

\begin{theorem}\label{coro:gain}
The matrix $ \L_{\t_0\btheta_0}$ in \eqref{Sigma} is $\0$ ($\bSigma_{\btheta_0} = \I_{\btheta_0}^{-1}$) if and only if 
\begin{equation}\label{cond}
  \mathrm{col}(\G_{\t_0\shared_0 }) \subseteq \mathrm{col}(\G_{\t_0\efree_0}),   
\end{equation}
where $\mathrm{col}(\B)$ is the space generated by columns of $\B$, 
and  $\shared$ and $\efree$ are the shared parameter and external free parameter given in Assumption \ref{connection}, and $\G_{\t_0\shared_0 }$ and $\G_{\t_0\efree_0}$ are given in \eqref{avar}. 
\end{theorem}

If \eqref{cond} occurs, then the external information is entirely absorbed by the estimation of $\efree$ without delivering any benefit to the estimation of $\btheta$. 
Obviously \eqref{cond} occurs 
in the extreme scenario where $\shared $ is empty
(there is no shared parameter) so that the primary and external sources are totally disconnected. The following discussion is about how to prevent \eqref{cond} when there is a shared parameter $\shared$ with $m>0$. 

Both $\G_{\t_0\shared_0 }$ and $\G_{\t_0\efree_0}$ have row dimension $H(K-1)$, 
where $H$ is the number of functions in the set ${\cal H}$ of functions we choose in \eqref{eq:Empirical_moments1}. The column dimensions of $\G_{\t_0\shared_0 }$ and $\G_{\t_0\efree_0}$ are respectively $m$ and $p(K-1)-m$,  the dimensions 
of $\shared$ and $\efree$ in  Assumption \ref{connection}, respectively. 
If ${\cal H}$ is chosen such that $ H(K-1) \leq p(K-1) -m$, then $\mathrm{col}\!\left(\G_{\t_0\efree_0}\right) $ has rank $H(K-1)$ and is in fact the entire $H(K-1)$-dimensional Euclidean space so that  \eqref{cond} holds regardless of what $\G_{\t_0\shared_0 }$ is. 
This means that
\begin{equation}\label{nec}
    H(K-1) > p(K-1)-m
\end{equation} 
is a necessary condition for \eqref{cond} not to hold.

Since the dimension of  $\mathrm{col}(\G_{\t_0\shared_0}) =  \min\{ m,  H(K-1)\}$ and the dimension of $\mathrm{col}(\G_{\t_0\efree_0})$ is $p(K-1)-m$ when \eqref{nec} holds, a sufficient condition for \eqref{cond} not to hold is 
\begin{equation}\label{suf}
   \min\{ m,  H(K-1)\} > p(K-1)-m ,
\end{equation} 
which is \eqref{nec}, adding that there are more shared parameters than external free parameters. 
In Example 1 of Section 2.3, there is no $\efree$ and $m = p(K-1)$ so that \eqref{suf} holds with $H=1$, that is, we can simply choose ${\cal H}$ having a constant function. 
In Example 2 of Section 2.3, 
$m = (p-1)(K-1)$ and \eqref{suf} holds if $p \geq 3$ and we choose an ${\cal H}$ with $H>1$. When $p=2$ in Example 2, \eqref{suf} cannot be achieved regardless of what $H$ is; but  \eqref{suf} is only sufficient (not necessary) to prevent \eqref{cond}.

In a given problem, we cannot choose $m$ since it is determined by the shared parameter $\shared$ in Assumption 2 and, thus, a general sufficient condition to prevent  \eqref{cond} is not available. 
What we can do is to enrich the set ${\cal H}$ so that at least \eqref{nec} holds. 
For example,  $H=1$ may be too small unless we are in the scenario of Example 1. 
Regardless of what $m$ is, the choice of $H \geq p$ ensures that the necessary condition \eqref{nec} holds. Since the amount of external information is fixed,  too large a ${\cal H}$ may not help and may in fact result in extra noise.

In the simulation study in Section 4, we  choose ${\cal H} $ as all components of $\Z$, in which case $H= $  the dimension of $\Z$.
Since we adopt the structure assumption in Example 2, this ${\cal H}$ ensures the sufficient condition \eqref{suf}. 

\subsection{Standard Errors}

To assess prediction error or make statistical inference on the target parameter $\btheta$, we need consistent  standard errors for the FMLE $\fmle$.
It suffices to consistently estimate the asymptotic covariance matrix  $\bSigma_{\btheta_0} $ in \eqref{Sigma}, using primary data. 
 In view of \eqref{Sigma} and the fact that $\J_{\t_0} = - \, \G_{\t_0\t_0}$ and $\I_{\btheta_0} = \G_{\btheta_0\btheta_0}$, the empirical Hessian 
 $\nabla^2_{\bgamma\bgamma}  \ell_n(\bgamma \mid \hat \q \,)\big|_{\bgamma=\hat\bgamma} $ can be used to consistently estimate $\bSigma_{\btheta_0} $, denoted by $\hat\bSigma_{\btheta} $. 

Although $\hat\bSigma_{\btheta} $ is consistent as $n \to \infty$, it may underestimate the sampling variability in finite samples and, thus, we follow the bootstrap alternative  \citep{efron1994introduction,shao2024gmm}. Specifically, we generate $ B$ bootstrap samples by sampling with replacement from the primary data $\{ (Y_i, \X_i), i=1,...,n\}$. For each bootstrap sample $b$, we compute the estimator in \eqref{hat.eata}, yielding $\hat{\bgamma}_b^*$, $b=1,...,B$. 
The bootstrap variance estimator for $\hat{\bgamma}$ is the sample covariance matrix of these $B$ bootstrap replicates $
\hat{\bgamma}_1^*, \ldots, \hat{\bgamma}_B^*$.

\subsection{Regularization for  Numerical Stability}
\label{sec:regularized}

Occasionally, directly maximizing \eqref{e1} can be numerically unstable because of the log-term in \eqref{e1}. For instance, if the $i$th log-term diverges to $-\, \infty$, then $\hat\delta_i$ diverges to $\infty$ (assigning essentially all the weight to observation $i$) and, thus, 
a maximizer of \eqref{e1} may not exist. In our experience, this instability arises when the Lagrange multiplier $\t$ moves too far away from a neighborhood of $\0$. To improve numerical stability and avoid such solutions, we want to search for a fused estimator by restricting $\t$ to remain near $\0$. Specifically, we replace \eqref{hat.eata} by the following $L_2$-penalized maximization, 
\begin{equation}\label{hat.eata.reg}
 \hat{\bgamma}  =
  \arg\max_{\bgamma}
  \left\{
    \ell_n(\bgamma \mid \hat\q)
    -
    \tau \|\t\|^2
  \right\},
\end{equation}
where $\tau > 0$ is a small regularization parameter. In our simulation studies, we set $\tau = 0.1$.

\section{Simulation} \label{sec:simulation}
 In this section, we conduct a Monte Carlo simulation to evaluate the finite-sample performance  
	of the proposed FMLE relative to the standard MLE that does not incorporate external machine-learning information.  
    
\subsection{Simulation Setting}\label{sec:sim-setting}

We consider a $K=3$ class multinomial setting \eqref{outcome-primary} with a  5-dimensional $\X$ ($p=5$), one primary study
($S=1$) with  sample size $n=500$, one external study ($S=0$) with sample size  $N=10{,}000$, and $n/N = 0.05$. 

We consider the proportion heterogeneity as in Example 2 of Section 2.3, where
the free parameters allow the two sources to differ in class prevalence through intercept shifts. Under this setting, condition \eqref{suf} is satisfied as long as $H>1$. The target parameters are $\btheta_1=(0.2, 1,-1,1,-1)^\top $ and $\btheta_2=(-0.1, -1,1,1,1)^\top $.
The free parameters (intercepts) are 
$\pfree = (0.2,\,-0.1)^\top$ and $\efree = (0.35,\,-0.25)^\top$ for internal and external sources, respectively.
The shared slope parameter $\shared $ contains the last 4 components of $\btheta_1$ and $\btheta_2$, $m=8$, and $\A_m $ is the identity. 

The primary study observes  $Y \in \{1,2,3\}$, while the external outcome is coarsened to a binary label
$U=1 \ \text{if } Y\in\{1,2\}$ and
$ U=2 \ \text{if } Y=3$.

The non-intercept components of $\X$ in the primary study have a 4-dimensional normal distribution with means 0, variances 1, and correlations 0.8. 
The corresponding covariate vector in the external source is generated from the 4-dimensional normal distribution with the following covariate shifts in mean and/or variance, but the same correlation of 0.8. \vspace{-2.5mm}
\begin{enumerate}
 \item   No shift: the external mean and variance remain the same as those in the primary \vspace{-3mm} study. 
 \item  Mean shift: the external mean is shifted to 
    $(0.06,-0.04,0.08,0)^\top$ but the external variance has no \vspace{-3mm} shift.
    \item Variance shift: the external variance is shifted to 2, but the external mean has no \vspace{-3mm} shift. 
    \item Mean and variance shift: both external mean and variance  are shifted according to the values in 2 and 3. \vspace{-3mm}
\end{enumerate}

 Two forms of external covariate $\Z$ are considered: 
 a full-feature setting in which $\Z = \X$ and 
 a missing-one-feature setting in which $\Z = \X_{{}_{-5}}$, ($\X$ without the 5th component). 
We choose $\mathcal{H}$ containing all components 
 of $\Z$,
with
 $H = \mbox{the dimension of } \Z $.

\begin{table}[tbp]
	\centering\begingroup\fontsize{14}{10}\selectfont
	\caption{Simulation bias, standard deviation (SD),  standard error (SE), and coverage probability (CP) of 95\% confidence intervals,
    based on 500 replications.}   
    \label{tbl:sim}
    \resizebox{\linewidth}{!}{
		\begin{tabular}[t]{llllrrrrrrrrrr}
			\toprule
			\multicolumn{4}{c}{ } & \multicolumn{5}{c}{$\btheta_1$} & \multicolumn{5}{c}{$\btheta_2$}  \\
			\cmidrule(l{3pt}r{3pt}){5-9} \cmidrule(l{3pt}r{3pt}){10-14} 
			$\Z$ & Shift & Metric & Method & \multicolumn{1}{c}{$0.2$} & \multicolumn{1}{c}{$1$} & \multicolumn{1}{c}{$-1$} & \multicolumn{1}{c}{$1$} & \multicolumn{1}{c}{$-1$} & \multicolumn{1}{c}{$-0.1$} & \multicolumn{1}{c}{$-1$} & \multicolumn{1}{c}{$1$} & \multicolumn{1}{c}{$1$} & \multicolumn{1}{c}{$1$} \\
			\midrule
			$ \X$
			& None
			& Bias & MLE & $-$0.010 & 0.025 & $-$0.025 & $-$0.011 & 0.003 & $-$0.012 & $-$0.010 & $-$0.005 & $-$0.007 & 0.041\\
			&  &  & FMLE  & $-$0.010 & 0.022 & $-$0.023 & $-$0.011 & 0.003 & $-$0.008 & 0.014 & 0.002 & $-$0.021 & 0.026\\
			\cmidrule(l){3-14}
			&  & SD & MLE & 0.143 & 0.255 & 0.268 & 0.241 & 0.252 & 0.156 & 0.305 & 0.300 & 0.289 & 0.288\\
			&  &  & FMLE  & 0.143 & 0.257 & 0.269 & 0.243 & 0.252 & 0.146 & 0.226 & 0.227 & 0.212 & 0.213\\
			\cmidrule(l){3-14}
			&  & SE & MLE & 0.141 & 0.258 & 0.258 & 0.250 & 0.257 & 0.158 & 0.296 & 0.296 & 0.281 & 0.296\\
			&  &  & FMLE & 0.184 & 0.276 & 0.276 & 0.263 & 0.275 & 0.173 & 0.223 & 0.227 & 0.241 & 0.222\\
			\cmidrule(l){3-14}
			&  & CP & MLE & 0.938 & 0.952 & 0.948 & 0.946 & 0.956 & 0.956 & 0.936 & 0.958 & 0.940 & 0.956\\
			&  &  & FMLE & 0.986 & 0.960 & 0.956 & 0.964 & 0.972 & 0.978 & 0.944 & 0.940 & 0.954 & 0.962\\
			\cmidrule(l){2-14}
			& Mean
			& Bias & MLE & 0.003 & 0.023 & $-$0.000 & 0.001 & $-$0.024 & $-$0.017 & $-$0.028 & 0.019 & 0.018 & 0.017\\
			&  &  & FMLE & 0.002 & 0.023 & $-$0.000 & $-$0.001 & $-$0.022 & $-$0.010 & 0.014 & 0.015 & $-$0.014 & 0.010\\
			\cmidrule(l){3-14}
			&  & SD & MLE & 0.149 & 0.259 & 0.253 & 0.256 & 0.250 & 0.169 & 0.304 & 0.314 & 0.292 & 0.283\\
			&  &  & FMLE & 0.149 & 0.260 & 0.254 & 0.258 & 0.251 & 0.152 & 0.216 & 0.224 & 0.231 & 0.213\\
			\cmidrule(l){3-14}
			&  & SE& MLE & 0.143 & 0.259 & 0.258 & 0.250 & 0.258 & 0.159 & 0.304 & 0.304 & 0.288 & 0.304\\
			&  &  & FMLE & 0.190 & 0.295 & 0.301 & 0.265 & 0.291 & 0.163 & 0.224 & 0.239 & 0.247 & 0.226\\
			\cmidrule(l){3-14}
			&  & CP & MLE & 0.944 & 0.958 & 0.954 & 0.948 & 0.966 & 0.934 & 0.952 & 0.936 & 0.956 & 0.964\\
			&  &  & FMLE & 0.984 & 0.974 & 0.972 & 0.954 & 0.978 & 0.962 & 0.944 & 0.936 & 0.934 & 0.952\\
			\cmidrule(l){2-14}
			& Variance
			& Bias & MLE & $-$0.010 & 0.025 & $-$0.025 & $-$0.011 & 0.003 & $-$0.012 & $-$0.010 & $-$0.005 & $-$0.007 & 0.041\\
			&  &  & FMLE & $-$0.011 & 0.023 & $-$0.024 & $-$0.013 & 0.004 & $-$0.016 & $-$0.023 & 0.015 & 0.011 & 0.040\\
			\cmidrule(l){3-14}
			&  & SD & MLE & 0.143 & 0.255 & 0.268 & 0.241 & 0.252 & 0.156 & 0.305 & 0.300 & 0.289 & 0.288\\
			&  &  & FMLE & 0.143 & 0.256 & 0.269 & 0.244 & 0.252 & 0.147 & 0.234 & 0.230 & 0.215 & 0.217\\
			\cmidrule(l){3-14}
			&  & SE & MLE & 0.141 & 0.258 & 0.258 & 0.250 & 0.257 & 0.158 & 0.296 & 0.296 & 0.281 & 0.296\\
			&  &  & FMLE & 0.184 & 0.276 & 0.275 & 0.262 & 0.275 & 0.172 & 0.228 & 0.229 & 0.239 & 0.223\\
			\cmidrule(l){3-14}
			&  & CP & MLE & 0.938 & 0.952 & 0.948 & 0.946 & 0.956 & 0.956 & 0.936 & 0.958 & 0.940 & 0.956\\
			&  &  & FMLE & 0.982 & 0.958 & 0.954 & 0.962 & 0.972 & 0.974 & 0.924 & 0.932 & 0.958 & 0.948\\
			\cmidrule(l){2-14}
			& Mean and
			& Bias & MLE & 0.003 & 0.023 & $-$0.000 & 0.001 & $-$0.024 & $-$0.017 & $-$0.028 & 0.019 & 0.018 & 0.017\\
			& variance &  & FMLE & 0.002 & 0.021 & $-$0.001 & $-$0.001 & $-$0.021 & $-$0.019 & $-$0.019 & 0.027 & 0.014 & 0.025\\
			\cmidrule(l){3-14}
			&  & SD & MLE & 0.149 & 0.259 & 0.253 & 0.256 & 0.250 & 0.169 & 0.304 & 0.314 & 0.292 & 0.283\\
			&  &  & FMLE & 0.149 & 0.261 & 0.253 & 0.260 & 0.251 & 0.151 & 0.228 & 0.228 & 0.228 & 0.220\\
			\cmidrule(l){3-14}
			&  & SE & MLE & 0.143 & 0.259 & 0.258 & 0.250 & 0.258 & 0.159 & 0.304 & 0.304 & 0.288 & 0.304\\
			&  &  & FMLE & 0.190 & 0.286 & 0.293 & 0.260 & 0.282 & 0.161 & 0.225 & 0.224 & 0.235 & 0.227\\
			\cmidrule(l){3-14}
			&  & CP &MLE & 0.944 & 0.958 & 0.954 & 0.948 & 0.966 & 0.934 & 0.952 & 0.936 & 0.956 & 0.964\\
			&  &  & FMLE & 0.984 & 0.976 & 0.968 & 0.956 & 0.978 & 0.950 & 0.936 & 0.928 & 0.962 & 0.936\\
			\midrule
			\addlinespace[0.3em]
			$\X_{{}_{-5}}$  & None
			& Bias & MLE & $-$0.010 & 0.025 & $-$0.025 & $-$0.011 & 0.003 & $-$0.012 & $-$0.010 & $-$0.005 & $-$0.007 & 0.041\\
			&  &  & FMLE & $-$0.010 & 0.022 & $-$0.022 & $-$0.011 & 0.002 & $-$0.011 & $-$0.003 & 0.001 & $-$0.016 & 0.040\\
			\cmidrule(l){3-14}
			&  & SD & MLE & 0.143 & 0.255 & 0.268 & 0.241 & 0.252 & 0.156 & 0.305 & 0.300 & 0.289 & 0.288\\
			&  &  & FMLE & 0.144 & 0.255 & 0.270 & 0.245 & 0.251 & 0.150 & 0.241 & 0.254 & 0.221 & 0.288\\
			\cmidrule(l){3-14}
			&  & SE & MLE & 0.141 & 0.258 & 0.258 & 0.250 & 0.257 & 0.158 & 0.296 & 0.296 & 0.281 & 0.296\\
			&  &  & FMLE & 0.184 & 0.277 & 0.278 & 0.266 & 0.272 & 0.182 & 0.245 & 0.258 & 0.247 & 0.321\\
			\cmidrule(l){3-14}
			&  & CP & MLE & 0.938 & 0.952 & 0.948 & 0.946 & 0.956 & 0.956 & 0.936 & 0.958 & 0.940 & 0.956\\
			&  &  & FMLE & 0.986 & 0.962 & 0.960 & 0.958 & 0.970 & 0.982 & 0.944 & 0.946 & 0.960 & 0.972\\
			\cmidrule(l){2-14}
			& Mean 
			& Bias & MLE & 0.003 & 0.023 & $-$0.000 & 0.001 & $-$0.024 & $-$0.017 & $-$0.028 & 0.019 & 0.018 & 0.017\\
			&  &  & FMLE & 0.003 & 0.021 & 0.003 & 0.000 & $-$0.024 & $-$0.012 & $-$0.001 & 0.014 & $-$0.005 & 0.017\\
			\cmidrule(l){3-14}
			&  & SD & MLE & 0.149 & 0.259 & 0.253 & 0.256 & 0.250 & 0.169 & 0.304 & 0.314 & 0.292 & 0.283\\
			&  &  & FMLE & 0.149 & 0.262 & 0.257 & 0.260 & 0.251 & 0.159 & 0.229 & 0.253 & 0.234 & 0.283\\
			\cmidrule(l){3-14}
			&  & SE & MLE & 0.143 & 0.259 & 0.258 & 0.250 & 0.258 & 0.159 & 0.304 & 0.304 & 0.288 & 0.304\\
			&  &  & FMLE & 0.192 & 0.302 & 0.322 & 0.273 & 0.279 & 0.180 & 0.253 & 0.285 & 0.284 & 0.326\\
			\cmidrule(l){3-14}
			&  & CP & MLE & 0.944 & 0.958 & 0.954 & 0.948 & 0.966 & 0.934 & 0.952 & 0.936 & 0.956 & 0.964\\
			&  &  & FMLE & 0.984 & 0.974 & 0.972 & 0.960 & 0.978 & 0.972 & 0.956 & 0.952 & 0.952 & 0.968\\
			\cmidrule(l){2-14}
			& Variance 
			& Bias & MLE & $-$0.010 & 0.025 & $-$0.025 & $-$0.011 & 0.003 & $-$0.012 & $-$0.010 & $-$0.005 & $-$0.007 & 0.041\\
			&  &  & FMLE & $-$0.010 & 0.022 & $-$0.021 & $-$0.010 & 0.001 & 0.022 & 0.053 & $-$0.124 & $-$0.063 & 0.039\\
			\cmidrule(l){3-14}
			&  & SD & MLE & 0.143 & 0.255 & 0.268 & 0.241 & 0.252 & 0.156 & 0.305 & 0.300 & 0.289 & 0.288\\
			&  &  & FMLE & 0.144 & 0.255 & 0.271 & 0.243 & 0.251 & 0.148 & 0.239 & 0.240 & 0.223 & 0.288\\
			\cmidrule(l){3-14}
			&  & SE & MLE & 0.141 & 0.258 & 0.258 & 0.250 & 0.257 & 0.158 & 0.296 & 0.296 & 0.281 & 0.296\\
			&  &  & FMLE & 0.188 & 0.285 & 0.291 & 0.271 & 0.272 & 0.184 & 0.246 & 0.265 & 0.282 & 0.323\\
			\cmidrule(l){3-14}
			&  & CP & MLE & 0.938 & 0.952 & 0.948 & 0.946 & 0.956 & 0.956 & 0.936 & 0.958 & 0.940 & 0.956\\
			&  &  & FMLE & 0.988 & 0.962 & 0.966 & 0.964 & 0.970 & 0.984 & 0.948 & 0.918 & 0.942 & 0.972\\
			\cmidrule(l){2-14}
			& Mean and
			& Bias & MLE & 0.003 & 0.023 & $-$0.000 & 0.001 & $-$0.024 & $-$0.017 & $-$0.028 & 0.019 & 0.018 & 0.017\\
			& variance &  & FMLE & 0.002 & 0.019 & 0.005 & $-$0.003 & $-$0.025 & 0.021 & 0.059 & $-$0.113 & $-$0.062 & 0.016\\
			\cmidrule(l){3-14}
			&  & SD & MLE & 0.149 & 0.259 & 0.253 & 0.256 & 0.250 & 0.169 & 0.304 & 0.314 & 0.292 & 0.283\\
			&  &  & FMLE & 0.149 & 0.262 & 0.259 & 0.262 & 0.250 & 0.153 & 0.225 & 0.244 & 0.239 & 0.283\\
			\cmidrule(l){3-14}
			&  & SE & MLE & 0.143 & 0.259 & 0.258 & 0.250 & 0.258 & 0.159 & 0.304 & 0.304 & 0.288 & 0.304\\
			&  &  & FMLE & 0.195 & 0.308 & 0.333 & 0.273 & 0.280 & 0.176 & 0.250 & 0.275 & 0.295 & 0.326\\
			\cmidrule(l){3-14}
			&  & CP & MLE & 0.944 & 0.958 & 0.954 & 0.948 & 0.966 & 0.934 & 0.952 & 0.936 & 0.956 & 0.964\\
			&  &  & FMLE & 0.984 & 0.972 & 0.964 & 0.948 & 0.980 & 0.974 & 0.950 & 0.932 & 0.922 & 0.968\\
			\bottomrule
	\end{tabular}
}
	\endgroup{}
\end{table}

\begin{table}[tbp]
\caption{The mean squared error of MLE and FMLE of class probabilities based on 500 replications.}
\label{tab:prob_classes}
\small
\centering
\begin{tabular}[t]{llrrrrrr}
\toprule
\multicolumn{2}{c}{ } & \multicolumn{3}{c}{$\Z = \X$} & \multicolumn{3}{c}{$\Z = \X_{{}_{-5}}$} \\
\cmidrule(l{3pt}r{3pt}){3-5} \cmidrule(l{3pt}r{3pt}){6-8}
Shift & Method & Class 1 & Class 2 & Class 3 & Class 1 & Class 2 & Class 3\\
\midrule
\multirow{2}{*}{None} & MLE & 0.0021 & 0.0021 & 0.0018 & 0.0021 & 0.0021 & 0.0018\\
 & FMLE & 0.0019 & 0.0019 & 0.0009 & 0.0019 & 0.0020 & 0.0012\\
\cmidrule(l){1-8}
\multirow{2}{*}{Mean} & MLE & 0.0020 & 0.0021 & 0.0017 & 0.0020 & 0.0021 & 0.0017\\
 & FMLE & 0.0019 & 0.0019 & 0.0009 & 0.0019 & 0.0020 & 0.0012\\
\cmidrule(l){1-8}
\multirow{2}{*}{Variance} & MLE & 0.0021 & 0.0021 & 0.0018 & 0.0021 & 0.0021 & 0.0018\\
 & FMLE & 0.0019 & 0.0019 & 0.0010 & 0.0020 & 0.0020 & 0.0014\\
\cmidrule(l){1-8}
\multirow{2}{*}{\shortstack[l]{Mean and\\Variance}} & MLE & 0.0020 & 0.0021 & 0.0017 & 0.0020 & 0.0021 & 0.0017\\
 & FMLE & 0.0019 & 0.0019 & 0.0009 & 0.0020 & 0.0021 & 0.0014\\
\bottomrule
\end{tabular}
\end{table}

\subsection{Results}
Based on 500 simulation replications, 
Table~\ref{tbl:sim} summarizes the empirical bias and standard deviation (SD) of the MLE $\mle$ and proposed FMLE $\fmle$ for target parameters, the standard error (SE) using the standard formula for the MLE and the bootstrap method described in Section 3.4 with $B=200$ for the FMLE, and the coverage probability (CP) of 95\% Wald confidence intervals for target parameters. 
For each simulation replication,  the MLE  $\mle$ is computed based on
primary data alone using the \texttt{multinom} function in \textsf{R}, and the proposed FMLE  $\fmle$ is computed using \eqref{hat.eata.reg} based on primary data and 
the external machine-learning prediction obtained by fitting an
XGBoost classifier to $(U, \Z)$. 

Across all simulation scenarios, both MLE and FMLE have negligible biases for all parameters.

The main advantage of the FMLE is its efficiency gain over the MLE for the regression coefficients in $\btheta_2$. Compared with the MLE, the SD is reduced by approximately 25\% in the full-feature setting ($\Z = \X$) and 15\% in the setting with one missing feature ($\Z = \X_{{}_{-5}})$. There is no gain for estimating  $\btheta_1$, which is expected because classes 1-2 are coarsened in the external source, and $\btheta_1$ represents the contrast between class 1 and class 2 under our model setting. On the other hand, the external information is  useful for estimating $\btheta_2$ representing the contrast between class 1 and class 3. The gains are stable across no shift, mean shift, variance shift, and mean plus variance shift, providing empirical evidence that the FMLE is robust to covariate shift, in agreement with the theoretical results in Section~\ref{sec:methodology}.

For confidence intervals, the CP related to the MLE is well calibrated, ranging from $0.934$ to $0.966$ across all parameters and scenarios. 
The CP related to the FMLE is likewise well calibrated except in a few cases 
for the coefficients in $\btheta_2$ corresponding to slope terms, where the CP decreases to approximately $0.92$. An explanation is that the uncertainty associated with external machine-learning prediction is not included in SE, which may sometimes have effects even with $n/N = 0.05$.

In addition to the performance of estimating the target $\btheta$, 
we also obtain the simulation mean squared error of the MLE and FMLE of each class probability $P (Y=k \mid S=1)$, a function of $\btheta$.  
The results are shown in Table \ref{tab:prob_classes}. 
Across all shift scenarios and feature sets, the FMLE uniformly outperforms the MLE. The improvement is particularly
  pronounced for class $k=3$, with a reduction of approximately 35\% compared to  7\% for the other two classes. This larger gain at class 3 is expected, as the external information is based on coarsened outcome labels for classes 1 and 2. 

 Overall, these findings suggest that the FMLE can substantially improve estimation efficiency for the parameters
most closely aligned with the external grouping structure, while maintaining generally satisfactory
interval coverage across a wide range of covariate shift scenarios.

\section{Real Data Example}\label{sec:Real_Data}

We illustrate the proposed fused estimation using data from the 2013-2018
cycles of the National Health and Nutrition Examination Survey (NHANES), a nationally representative, repeated cross-sectional survey conducted by the U.S.\
Centers for Disease Control and Prevention.

\subsection{Data Sources and Heterogeneity}

The primary study consists of 9,186 sampled units that completed blood pressure tests and fasting
laboratory examinations. The outcome $Y$ in our analysis is the blood pressure classification according to the following three categories:
\begin{center}
\small
\begin{tabular}{lcccc}
\toprule
&  & Systolic (mmHg) & & Diastolic (mmHg) \\ \cline{3-5}
Normal & $Y=1$ &  $< 130$ & and & $ < 85$  \\
Prehypertension & $Y=2$ &  130--140  & or & 85--90 \\
Hypertension & $Y=3$ & $>140$ & or& $>90$ \\
\bottomrule
\end{tabular}
\end{center}
Each unit in the  primary study has 14 covariates: 8 demographic and anthropometric variables age, sex, race, income-to-poverty ratio (income), body mass index (BMI), waist circumference (waist), height, and weight, and 6  laboratory results 
glucose, insulin, triglycerides (TG), low-density lipoprotein cholesterol (LDL), high-density lipoprotein cholesterol (HDL), and total cholesterol (TC).

The NHANES contains another dataset of 12,425 sampled units that do not have laboratory examination but have 8 demographic and anthropometric covariates and blood pressure results. This less informative dataset is used as the external source for data fusion.

\begin{figure}[tbp]
  \centering
  \includegraphics[width=0.5\linewidth]{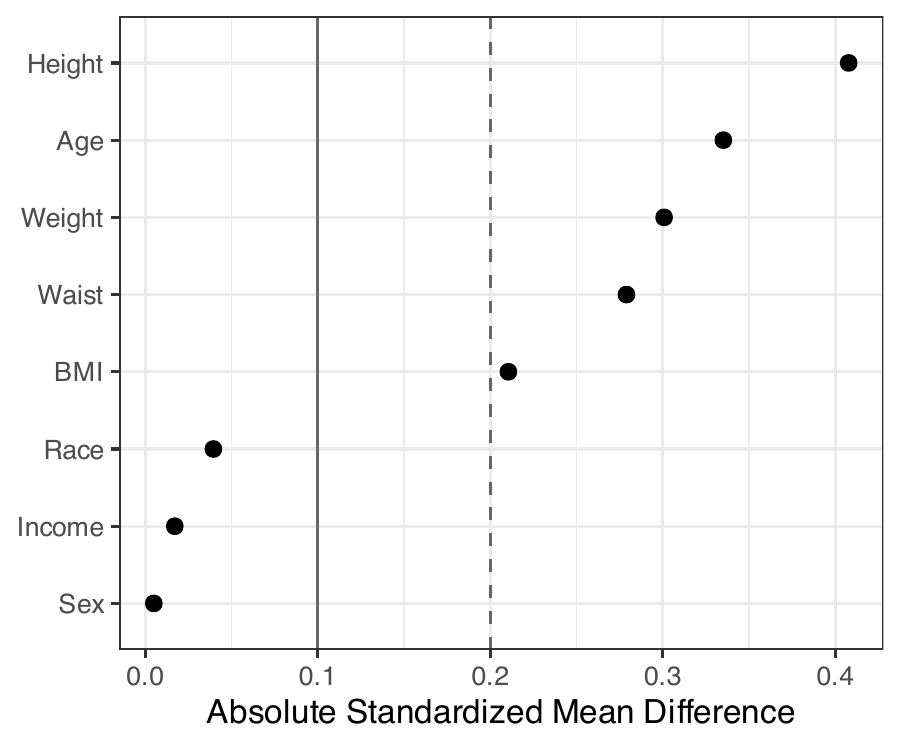}
  \caption{Standardized mean differences for shared covariates between the primary and external sources.}
  \label{fig:love}
\end{figure}

We now examine two types of heterogeneity between the primary and
external sources, as we discussed in Section 2.3.  
First, Figure~\ref{fig:love} presents a 
plot displaying absolute standardized mean differences of the 8 shared demographic and anthropometric covariates across the primary and external sources.
Age, height, weight, waist,  and BMI exhibit substantial discrepancies, 
using the common threshold of $0.2$, which indicates a substantial covariate shift.
Second, 
Table~\ref{tbl:rate} lists the empirical class proportions for blood pressure in the primary and external samples. 
The proportion of normal blood pressure is notably higher in the external source, whereas the primary source exhibits higher proportions of both prehypertension and
hypertension. These differences reflect heterogeneity (concept shifts) in 
outcome prevalence between the primary and external sources. 

\begin{table}[tbp]
\centering
\caption{Empirical class proportions of blood pressure categories in the two samples.}
\label{tbl:rate}
\small
\renewcommand{\arraystretch}{1.1}
\begin{tabular}{lcc}
\toprule
Category & Primary & External \\
\midrule
Normal           & 0.704 & 0.754 \\
Prehypertension  & 0.153 & 0.134 \\
Hypertension     & 0.146 & 0.112 \\
\bottomrule
\end{tabular}
\end{table}

\subsection{Standard and Fused Estimation}

Estimation of parameters in \eqref{outcome-primary} using data only from 9,186 units in the primary study is standard by maximizing likelihood \eqref{primary-likelihood}. To see if we can use external information to gain  efficiency, 
the fused estimation is designed to adjust for
covariate shift and outcome-generating heterogeneity demonstrated in  Figure~\ref{fig:love} and
Table~\ref{tbl:rate}, and is applied using likelihood \eqref{e1},  
under the shared parameter structure in Example 2 to connect the two sources. 

The external information consists of a machine-learning prediction using XGBoost based on data from all 12,425 external units with 3 blood pressure categories and 8 shared demographic and anthropometric covariates. To avoid overfitting, the external sample is randomly split into training and validation sets in a $4{:}1$ ratio, and early stopping based on validation loss is employed.
To apply the proposed fused estimator, we adopt  ${\cal H} = $ all 8  demographic and anthropometric covariates plus an intercept. 

\begin{figure}[tbp]
    \centering
    \includegraphics[width=1\linewidth]{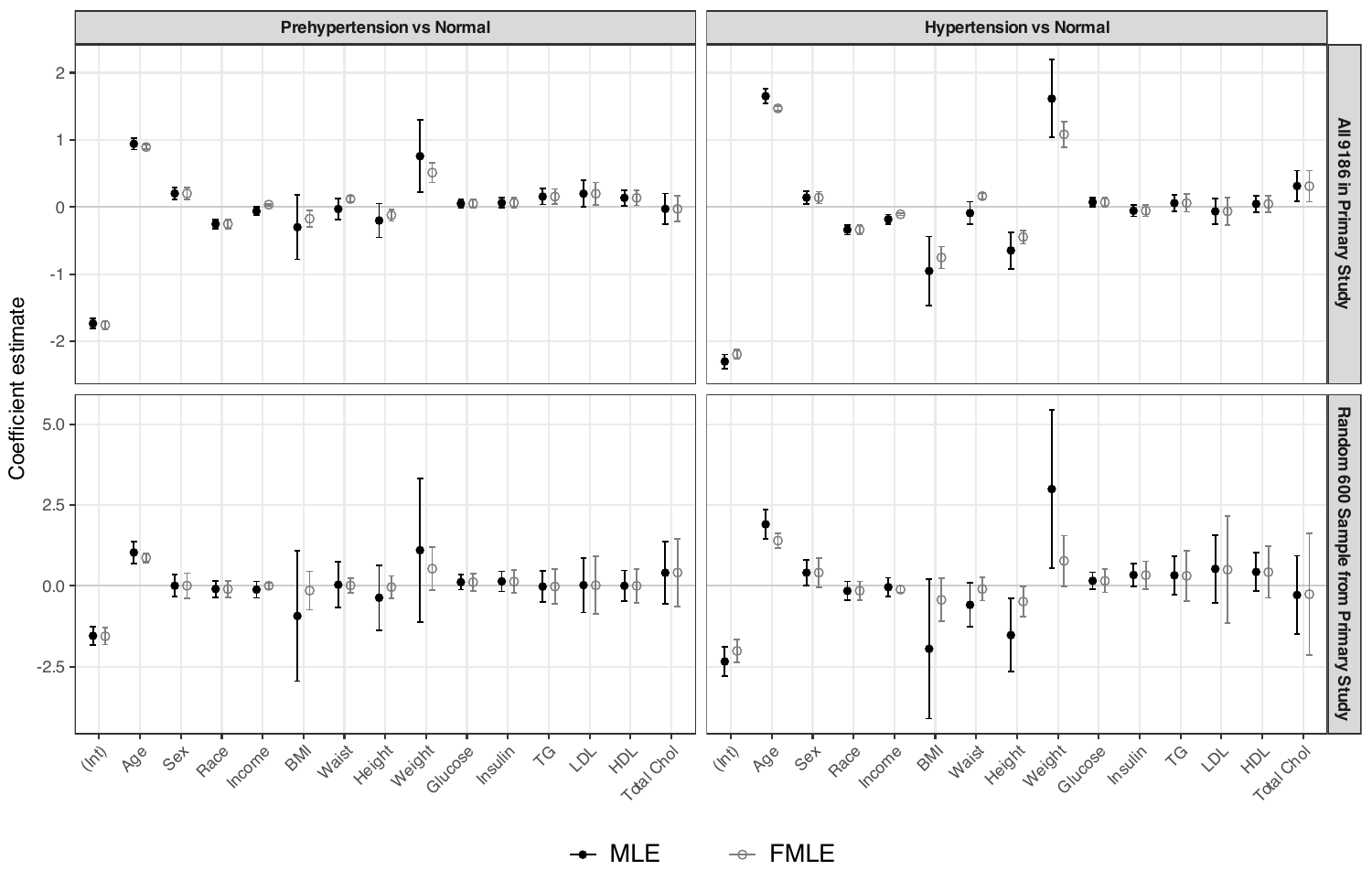}
     \caption{Point estimates (dots) and confidence intervals (vertical bars).}
    \label{fig:real}
\end{figure}

Based on model \eqref{outcome-primary}, 
estimates of  $\btheta_1$ (corresponding to prehypertension versus normal)  and $\btheta_2$ (corresponding to hypertension versus normal) broken down to each covariate component, and the associated 95\% Wald confidence intervals are shown in the top two panels of Figure \ref{fig:real}, where the MLE using primary data alone are presented with solid dots and the proposed FMLE are presented with circles.  
Since the external source does not have 6 laboratory covariates, as expected, the results from the two estimation methods for these covariates are about the same. 
For 8 demographic and anthropometric covariates, the proposed FMLE has some improvement over the standard MLE using primary data alone, where improvements for height, weight, waist, and BMI are appreciable.

The ratio $n/N$ in this example is $9,186/12,425 \approx 0.74$. 
 To see the effect with a smaller sample size ratio $n/N$, 
we create a random sample of size 600 (without replacement) from the primary dataset of 9,186 units, and
 treat this random sample as the primary dataset to compute the MLE and FMLE and their associated confidence intervals, where
 fused estimation uses the same external information from 12,425 units. In this way, the ratio $n/N$ becomes $600/12,425 \approx 0.048$, close to 0.05 in the simulation (Section 4). 
 The results are shown in the bottom panels of Figure \ref{fig:real}. 

It can be seen from Figure \ref{fig:real} that the point estimates are comparable for the two cases with primary sample sizes 600 and 9,186, but the confidence intervals based on MLE with 600 sample size are much wider, and the fused FMLE provides substantially tighter confidence intervals for all 8 demographic and anthropometric covariates.
The results show that fused analysis is more useful in the case where the ratio $n/N$ is smaller.

\section{Discussion}\label{sec:Discussion}

This paper proposes data fusion for multiclass classification
that leverages a robust and efficient machine-learning prediction rule constructed with a large external dataset to improve parametric multinomial logistic regression in a primary study with a much smaller size, without requiring individual-level external data. 
The proposed fused  estimators accommodate several practical challenges simultaneously:
partial covariates, coarsened class
labels, and heterogeneity in both covariate distributions and outcome-generating mechanisms between the external and primary populations. 
 We establish
consistency, asymptotic normality, and efficiency properties of the fused
estimators, and confirm the theoretical gains through simulation studies and a real data application in the NHANES. 
Our results demonstrate that carefully integrating external machine-learning predictions into a primary likelihood-based analysis with heterogeneity appropriately handled can substantially improve efficiency.
A key issue of the proposed framework is the validity of the moment condition
\eqref{eq:adv+}, which relies on Assumption 3 when the external source does not have all covariates in $\X$, and the validity of the likelihood  
\eqref{likelihood}, which relies on the structural Assumption \ref{connection} connecting two sources of data. 
If either of these assumptions is violated,  then some moment constraints in \eqref{eq:Empirical_moments1} may not be valid and the resulting fused estimator may be biased.
A possible remedy is  to consider a data-driven shrinkage approach that
filters out invalid constraints.
A detailed study of such an approach and its theoretical properties is left for future work.

Several other extensions remain open for further research, including
high-dimensional versions of the fused estimator,
robustness against misspecification of the primary model, parametric models other than multinomial logistic regression,
multiple external studies, each contributing
its own prediction rule, 
and more complex
outcomes such as longitudinal responses, survival data, functional predictors, or
multi-stage and ordinal outcomes. 

\acks{Chi-Shian Dai's research was partially supported by NSTC of Taiwan under Grant 114-2118-M-006-001-MY2.}

\appendix
\section{Regularity Conditions and Proofs}\label{sec:proof}

Our proofs follow standard arguments for M-estimation. Let
$\ell_n(\bgamma\mid\q \, )$ be $\ell_n(\bgamma\mid \hat\q \, )$ with $\hat \q$ replaced by $\q$, 
where both $\hat \q$ and $\q$ are given in Section  \ref{sec:observed}.
\subsection{Regularity Conditions for Theorem \ref{thm:consistency}}

The following regularity conditions are needed for the consistency of $\hat \bgamma$ in Theorem \ref{thm:consistency}. \vspace{-2mm}

\begin{itemize}
  \item[(C1)] 
 The parameter space $\bGamma$ of $\bgamma $ is a compact set and  $E \{ \ell_1 (\bgamma \mid \q \, )\} $ has a unique maximizer
  at $\bgamma_0  \in \bGamma$. \vspace{-2mm}
  \item[(C2)] The class of functions, $E \{ \ell_1 (\bgamma \mid \q \, )\}$,  $\bgamma\in\bGamma $, 
  is Glivenko-Cantelli \citep{geer2000empirical}. \vspace{-2mm}
  \item[(C3)] There exists a neighborhood $\mathcal{N}$ of $\q$ such that\vspace{-3mm}
  \[
  \sup_{\bgamma\in\bGamma ,\,\q\in\mathcal{N}}
  \left|
  1+\sum_{l=1}^{L-1}\sum_{h\in\mathcal{H}} g_{l,h} (\X \mid \bphi , \q )
 \, \lambda_{l,h}
    \right|^{-1} \leq c_0 \qquad \mbox{almost surely,} 
  \]
  where $g_{l,h} (\X \mid \bphi , \q ) = \sum_{k \in {\cal C}_l} \{ p_k(\X \mid \bphi )  - q _k(\Z )\}h (\Z )$ and $c_0$ is a positive constant.  
\end{itemize}
\subsection{Proof of Theorem \ref{thm:consistency}}
Note that
\begin{equation}\label{r1}
\sup_{\bgamma\in\bGamma}
\bigl|\ell_n(\bgamma\mid\hat{\q}\, )- E \{ \ell_1 (\bgamma \mid \q \, )\} \bigr|
\le
\sup_{\bgamma\in\bGamma}
\bigl|\ell_n(\bgamma\mid\hat{\q})-\ell_n(\bgamma\mid\q)\bigr|
 +
\sup_{\bgamma\in\bGamma}
\bigl|\ell_n(\bgamma\mid\q)-E \{ \ell_1 (\bgamma \mid \q \, )\}\bigr|.
\end{equation}
(C2) implies that the second term on the right side of \eqref{r1} $\to 0$ almost surely 
\citep{geer2000empirical}.
By Assumption 1(i), $\|\hat{\q}-\q\|_\infty \xrightarrow{p} 0$ and, hence, 
 with probability tending to one, $\hat{\q}$ lies in the neighborhood
$\mathcal{N}$ of $\q$ in (C3). Thus, 
the first term on the right side of \eqref{r1} is bounded by
\begin{align*}
& \
\frac{1}{n}\sum_{i=1}^n
\sup_{\bgamma\in\bGamma}
\left|
\log\!\left(
1+\sum_{l=1}^{L-1}\sum_{h\in\mathcal{H}}
g_{l,h} (\X_i \mid \bphi , \hat \q )  \,\lambda_{l,h}
\right)
-
\log\!\left(
1+\sum_{l=1}^{L-1}\sum_{h\in\mathcal{H}}
g_{l,h} (\X_i \mid \bphi , \q ) \,\lambda_{l,h}
\right)
\right| \\
\le & \
\frac{1}{n}\sum_{i=1}^n
\sup_{\bgamma\in\bGamma,\,\q\in\mathcal{N}}
\left|
1 \! + \! \sum_{l=1}^{L-1}\sum_{h\in\mathcal{H}}
g_{l,h}\bigl(\X_i\mid \bphi, \q \bigr)\,\lambda_{l,h}
\right|^{-1} \!\!
\sup_{\bgamma\in\bGamma}
\left|
\sum_{l=1}^{L-1}\sum_{h\in\mathcal{H}}
\{\hat q_k(\Z_i)-q_k(\Z_i)\}\,h(\Z_i)\,\lambda_{l,h}
\right| \\
\le & \ c_0 \|\hat{\q}-\q \|_\infty \ \frac{1}{n}\sum_{i=1}^n \sup_{\bgamma\in\bGamma} \sum_{l=1}^{L-1}\sum_{h\in\mathcal{H}} | h(\Z_i)| \, | \lambda_{l,h}| \\
 \xrightarrow{p} & \ 0 , 
\end{align*}
where the first inequality follows from 
 a first-order Taylor expansion of the logarithm, the second inequality follows from (C3), and  $\xrightarrow{p} 0$ follows from Assumption 1(i) and the fact that $h(\Z)$ is integrable. 
Therefore, the first term on the right side of \eqref{r1} $\xrightarrow{p} 0$.  
This shows that the left side of \eqref{r1} $\xrightarrow{p} 0$. 
This  uniform convergence together with (C1) imply consistency of the M-estimator, that is, $\hat{\bgamma} \xrightarrow{p} \bgamma_0$ 
\citep[Theorem~2.1]{newey1994large}. 

\subsection{Regularity Conditions for Theorem \ref{thm:clt.regu}}

In addition to (C1)-(C3), the following regularity conditions are needed for the asymptotic normality of $\hat \bgamma$ in Theorem \ref{thm:clt.regu}.

\begin{itemize}
 \item[(C4)]
  $\E\bigl\{\nabla_{\bgamma}\ell_1( \bgamma \mid \q )\big|_{\bgamma =\bgamma_0} \bigr\}=\mathbf{0}$ and 
  $\var \{ \nabla_{\bgamma}\ell_1 (\bgamma \mid \q )\big|_{\bgamma =\bgamma_0} \} $ is finitely defined. 

  \item[(C5)] 
 With $\| \A \|_{\mathrm{op}} = $ the maximum of absolute values of eigenvalues of the matrix $\A$,  
    \[
    \E \Bigl\{ \sup_{\bgamma \in \mathcal{M}} \bigl\|
        \nabla^2_{\bgamma\bgamma} \ell_n (\bgamma \mid   \q )\big|_{\bgamma =\bgamma_0}
      \bigr\|_{\mathrm{op}}
    \Bigr\} < \infty, 
  \]
  and 
  there exist a measurable function $b(\X,Y)$, a constant $\epsilon > 0$, and a
  neighborhood $\mathcal{M}$ of $\bgamma_0$ such that for $\hat \q$ with small enough $\| \hat \q - \q \|_{\infty}$, 
  $$
    \sup_{\bgamma \in \mathcal{M}}
    \bigl\|
      \nabla^2_{\bgamma\bgamma} \ell_1(\bgamma \mid \hat \q)
      - \nabla^2_{\bgamma\bgamma} \ell_1(\bgamma  \mid  \q)
    \bigr\|_{\mathrm{op}}
    \le b(\X,Y) \, \| \hat \q - \q\|_\infty^{\epsilon} . $$
and $E\{b(\X,Y)\} \, \| \hat \q - \q\|_\infty^{\epsilon}\xrightarrow{p}0$.
 
  \item[(C6)] The matrix $\G_{\bgamma_0}$ in \eqref{avar} is non-singular.
\end{itemize}
\subsection{Proof of Theorem \ref{thm:clt.regu}}
When $\bgamma$ is $\bgamma_0$, the corresponding $\t_0 = \mathbf{0}$, 
and $\btheta $ and $\efree$ are denoted by $ \btheta_0$ and $\efree_0$. 
A direct calculation shows that
$\nabla_{\btheta}\ell_1( \bgamma \mid \q \, )\big|_{\bgamma= \bgamma_0}
= \nabla_{\btheta}\ell_1(\btheta ) \big|_{\btheta= \btheta_0}$
and $\nabla_{\efree}\ell_1(\bgamma \mid \q \, )\big|_{\bgamma= \bgamma_0}
=\mathbf{0}$. Also, 
$\nabla_{\btheta}\ell_1(\btheta ) \big|_{\btheta= \btheta_0}$
has mean $\mathbf{0}$ under (C4) and is uncorrelated with $\nabla_{\t}\ell_n( \bgamma \mid \q \, )\big|_{\bgamma= \bgamma_0}$.
With $\J_{\btheta_0} = \var \{ \nabla_{\btheta}\ell_1(\btheta ) \big|_{\btheta= \btheta_0}\}$ and  $\J_{\t_0} = \var \{ \nabla_{\t} \ell ( \bgamma \mid \q \, ) \big|_{\bgamma = \bgamma_0}\}$, 
we obtain that under (C4), 
\begin{equation}
     \sqrt{n} \,
  \nabla_{\bgamma}\ell_n (\bgamma \mid \q \, ) \, \big|_{\bgamma = \bgamma_0}
  \;\xrightarrow{d}\;
  N(\mathbf{0},\J_{\bgamma_0}) \label{r2}
\end{equation}
with $\J_{\bgamma_0}$ given by \eqref{avar}. 
By the definition of $\overline{\q}$,  
\begin{align}\label{r3}
&  \ n \left\{ 
 \nabla_{\t}\ell_n (\bgamma \mid \overline{\q} \, ) \big|_{\bgamma= \bgamma_0}-
\nabla_{\t}\ell_n (\bgamma \mid \q \, )\big|_{\bgamma= \bgamma_0} \right\}^2\notag\\
=& \ \frac{1}{n} \left[  \sum_{i=1}^n \sum_{l=1}^{L-1} 
\sum_{h \in {\cal H}} \{\overline{q}_l (\Z_i ) - q_l(\Z_i) \} h(\Z_i) \right]^2  \notag \\
 \leq & \ \frac{1}{n}\sum_{i=1}^n  \sum_{l=1}^{L-1}\sum_{h\in\mathcal{H}} \{\overline{q}_l (\Z_i ) - q_l (\Z_i)\}^2 \sum_{i=1}^n  \sum_{l=1}^{L-1}\sum_{h\in\mathcal{H}} h^2(\Z_i) .
\end{align}
Note that
 \[\sum_{i=1}^n  \sum_{l=1}^{L-1}\sum_{h\in\mathcal{H}} \{\overline{q}_l (\Z_i ) - q_l (\Z_i)\}^2  \leq 
 (L-1)H \, n \, \| \overline{\q} - \q \|_\infty^2   \] 
 and 
 \[ E\left[ \sum_{i=1}^n  \sum_{l=1}^{L-1}\sum_{h\in\mathcal{H}} \{\overline{q}_l (\Z_i ) - q_l (\Z_i)\}^2 \right] =  Hn E \| \overline{\q}(\Z) - \q (\Z) \|^2 . \]
In any case, 
it follows from Assumption 1(ii) that 
 \[\sum_{i=1}^n  \sum_{l=1}^{L-1}\sum_{h\in\mathcal{H}} \{\overline{q}_l (\Z_i ) - q_l (\Z_i)\}^2  \, \xrightarrow{p} \, 0 .  \] 
 Since $h^2(\Z)$ is integrable, this shows that 
 \eqref{r3} 
$ \, \xrightarrow{p} \, 0 $ 
and hence \eqref{r2} holds with $\q$ replaced by $\overline{\q}$. 
Define
$\delta_{l h}(\Z) = \sum_{k \in {\cal C}_l}
\bigl[\hat q_k(\Z)-\E\bigl\{\hat q_k(\Z)\mid\Z\bigr\}\bigr]\,h(\Z)$,
$l=1,\ldots,L-1$, $h\in\mathcal{H}$.
Then 
$
\E\bigl[\delta_{l h}(\Z)\mid \Z\bigr]=0$, 
$\var\bigl\{\delta_{l h}(\Z)\mid \Z\bigr\}
=
\var\bigl\{\sum_{k \in {\cal C}_l}\hat q_k(\Z)\mid \Z\bigr\}\,\{h(\Z)\}^2
=
\sigma^2_l(\Z)\, h^2(\Z)$, 
\[  \sqrt{n} \,
\nabla_{\t}\ell_n (\bgamma \mid \hat\q \, ) \big|_{\bgamma= \bgamma_0}  = 
\sqrt{n} \,\nabla_{\t}\ell_n (\bgamma \mid \overline\q \, )\big|_{\bgamma= \bgamma_0}  
+ \frac{1}{\sqrt{n}} \sum_{i=1}^n \sum_{l=1}^{L-1} \sum_{h \in {\cal H}}  \bdelta_{l h} (\Z_i) , \]
and
\[ \var \{  \sqrt{n} \,
\nabla_{\t}\ell_n (\bgamma \mid \hat\q \, ) \big|_{\bgamma= \bgamma_0} \} = 
\var \{ \sqrt{n} \,\nabla_{\t}\ell_n (\bgamma \mid \overline\q \, )\big|_{\bgamma= \bgamma_0}  \}
+ \sum_{l=1}^{L-1} \sum_{h \in {\cal H}}  E \{ \sigma^2_l (\Z) h^2 (\Z) \} , \]
where the last term $\to 0$ under Assumption 1(iii). This shows that \eqref{r2} holds with $\q$ replaced by $\hat{\q}$, that is, under Assumption 1, the contribution of the external prediction noise to the asymptotic covariance of the score function is asymptotically negligible. 

To complete the proof of the asymptotic normality of $\hat\bgamma$, we use \eqref{asy}, which is obtained from a standard Taylor expansion, and establish  the convergence of the empirical Hessian, 
\[ \nabla^2_{\bgamma\bgamma} \ell_n ( \bgamma \mid \hat\q \,) \big|_{\bgamma = \bgamma_0} \, \xrightarrow{p} \, \G_{\bgamma_0} 
\]
with $\G_{\bgamma_0}$ given by \eqref{avar}, under (C5) \citep[Theorem 8.2]{newey1994large}. The existence of 
$\G_{\bgamma_0}^{-1}$ is assumed under (C6). 

The block form of $\G_{\bgamma_0}$ can be easily verified using the fact that $\t_0 = \mathbf{0}$. The fact that $\J_{\btheta_0} = \G_{\btheta_0\btheta_0}$ follows from the standard analysis.
Finally, we show that $\J_{\t_0} = - \G_{\t_0\t_0}$. 
Note that 
\[\nabla_{\lambda_{l,h}}\ell(\gamma\mid \q)\big|_{\bgamma = \bgamma_0}=- \, g_{l,h}(\X \mid \A_m \shared_0 , \efree_0 )\]
and 
\[\nabla_{\lambda_{l,h},\lambda_{l',h'}}\ell(\gamma\mid \q)\big|_{\bgamma = \bgamma_0}= g_{l,h}(\X \mid \A_m \shared_0 , \efree_0 )g_{l',h'}(\X \mid \A_m \shared_0 , \efree_0 ).\]
As a result, 
$$\nabla_{\lambda_{l,h}}\ell(\gamma\mid \q)\nabla_{\lambda_{l',h'}}\ell(\gamma\mid \q)\big|_{\bgamma = \bgamma_0}=\nabla_{\lambda_{l,h},\lambda_{l',h'}}\ell(\gamma\mid \q)\big|_{\bgamma = \bgamma_0} $$
and $\J_{\t_0}=-\G_{\t_0\t_0}$ follows from the definitions of $\J_{\t_0}$ and $ \G_{\t_0\t_0}$, and the fact that $E\{g_{l,h}(\X \mid \A_m \shared_0 , \efree_0 )\}=0$.

\subsection{Proof of Theorem \ref{thm:eff1}}
Let $\ba=(\t^\top,\efree^\top)^\top$ denote the parameters other than $\btheta$ and 
$\ba_0$ be $\ba$ when $\bgamma = \bgamma_0$, and 
let $\B_{\btheta_0}$ and $\B_{\btheta_0\ba_0} $ be respectively the $\btheta_0\times \btheta_0$-block and $\btheta_0\times \ba_0$-block of $\G^{-1}_{\bgamma_0}$. By the block inverse formula for partitioned matrices, 
$\B_{\btheta_0}   = \I_{\btheta_0}^{-1}
    + \I_{\btheta_0}^{-1}\,
      \G_{\btheta_0\ba_0}\,
      \bm{S}_{\ba_0}^{-1}\,
      \G_{\ba_0\btheta_0}\,
      \I_{\btheta_0}^{-1}$
and $  \B_{\btheta_0\ba_0}
  = - \, \I_{\btheta_0}^{-1}\,
      \G_{\btheta_0\ba_0}\,
      \bm{S}_{\ba_0}^{-1}$, where 
      \begin{equation}\label{r4}
          \bm{S}_{\ba_0}= \G_{\ba_0\ba_0}
- \G_{\ba_0\btheta_0}\,
\I_{\btheta_0}^{-1}\,
\G_{\btheta_0\ba_0}=\begin{pmatrix}
\D_{\t_0}  & \G_{\t_0\efree_0}\\
\G_{\efree_0\t_0}    &\0
\end{pmatrix}. 
      \end{equation}
Note that $\bSigma_{\btheta_0}$ is the $\btheta_0 \times \btheta_0$–block of
$\G_{\bgamma_0}^{-1}\J_{\bgamma_0}\G_{\bgamma_0}^{-1}$. 
By the block inverse formula for partitioned matrices,
\begin{align*}
 \bSigma_{\btheta_0}
  &= \B_{\btheta_0}
     \J_{\btheta_0}\,
    \B_{\btheta_0}
   + \B_{\btheta_0\ba_0}\begin{pmatrix} \J_{\t_0} & \0 \\
   \0 & \0 \end{pmatrix} 
   \B_{\btheta_0\ba_0}^\top  \\
  &= \I_{\btheta_0}^{-1}
     + \I_{\btheta_0}^{-1}\,
       \G_{\btheta_0\ba_0}\,
       \S_{\ba_0}^{-1}\,
       \begin{pmatrix}
         \D_{\t_0} & 2\G_{\t_0\efree_0}\\
          2\G_{\efree_0\t_0} &\0
       \end{pmatrix}\,
       \S_{\ba_0}^{-1}\,
       \G_{\ba_0\btheta_0}\,
       \I_{\btheta_0}^{-1} \\
      &= \I_{\btheta_0}^{-1} +
        \begin{pmatrix} \L \\ \C \end{pmatrix}^\top 
       \begin{pmatrix}
         \D_{\t_0} & 2\G_{\t_0\efree_0}\\
          2\G_{\efree_0\t_0} &\0
       \end{pmatrix}\,
        \begin{pmatrix} \L \\ \C \end{pmatrix} , 
\end{align*}
where 
\begin{equation}\label{r5}
 \begin{pmatrix} \L \\ \C \end{pmatrix} = \bm{S}_{\ba_0}^{-1}\G_{\ba_0\btheta_0}\I_{\btheta_0}^{-1}= \bm{S}_{\ba_0}^{-1}\begin{pmatrix} \G_{\t_0\btheta_0}\I_{\btheta_0}^{-1}\\ \mathbf{0} \end{pmatrix},   
\end{equation}  
and we used $\J_{\btheta_0} = \I_{\btheta_0} = \G_{\btheta_0\btheta_0}$, 
$\J_{\t_0} = - \, \G_{\t_0\t_0}$, and $\G_{\ba_0\btheta_0} ={ \G_{\t_0\btheta_0} \choose \0 }$.   From \eqref{r4}-\eqref{r5}, 
\[ \begin{pmatrix} \G_{\t_0\btheta_0}\I_{\btheta_0}^{-1}\\ \mathbf{0} \end{pmatrix} =  \bm{S}_{\ba_0} \begin{pmatrix} \L \\ \C \end{pmatrix} =  \begin{pmatrix}
\D_{\t_0}  & \G_{\t_0\efree_0}\\
\G_{\efree_0\t_0}    &\0
\end{pmatrix}  \begin{pmatrix} \L \\ \C \end{pmatrix}, \]
which implies that we must have $\G_{\efree_0\t_0}  \L = \0$. Therefore, 
\[ \bSigma_{\btheta_0} = \I_{
\btheta_0}^{-1} + \L^\top \D_{\t_0} \L .  \]
From \eqref{r5}, $\L = \B_{\t_0}\G_{\t_0\btheta_0} \I_{\btheta_0}^{-1} $, where $\B_{\t_0}$ is the $\t_0 \times \t_0$-block of $\S_{\ba_0}^{-1}$. By the block inverse formula for $\S_{\ba_0}$, 
$$\B_{\t_0}=\D_{\t_0}^{-1}-\D_{\t_0}^{-1}\G_{\t_0\efree_0}\bigl(\G_{\efree_0\t_0}\D_{\t_0}^{-1}\G_{\t_0\efree_0}\bigr)^{-1}\G_{\efree_0\t_0}\D_{\t_0}^{-1} . $$
This proves that $\L = \L_{\t_0\btheta_0}$ given in \eqref{Sigma}. The proof of Theorem \ref{thm:eff1} is completed because \eqref{eq:eff} follows from \eqref{Sigma} as $\D_{\t_0}$ is negative definite.

\subsection{Proof of Theorem \ref{coro:gain}}
It suffices to show that $\L_{\t_0\btheta_0} = \0$ is equivalent to \eqref{cond}. 
Since $\D_{\t_0}$ is  negative definite, there exists a symmetric positive definite matrix $\H$ such that $\D_{\t_0}^{-1}=-\H^2$. Thus, 
\begin{align*}
    \L_{\t_0\btheta_0} &  = \{\D_{\t_0}^{-1}-\D_{\t_0}^{-1}\G_{\t_0\efree_0}\bigl(\G_{\efree_0\t_0}\D_{\t_0}^{-1}\G_{\t_0\efree_0}\bigr)^{-1}\G_{\efree_0\t_0}\D_{\t_0}^{-1} \} \G_{\t_0\btheta_0} \I_{\btheta_0}^{-1} \\
    & =-\H\{\I-\R(\R^\top\R)^{-1}\R^\top\}\H\,\G_{\t_0\btheta_0}\I_{\btheta_0}^{-1} ,
\end{align*}
where  $\R=\H\G_{\t_0\efree_0}$ and
$\I$ is the identity matrix of dimension $=$ the dimension of $\t$.
Consequently,
 $\L_{\t_0\btheta_0} = \0$  if and only if
$
\{\I-\R(\R^\top\R)^{-1}\R^\top\}\H\,\G_{\t_0\btheta_0}=\0$. 
The matrix $\R(\R^\top\R)^{-1}\R^\top$ is the projection onto the column space of $\R=\H\G_{\t_0\efree_0}$. Hence, $\L_{\t_0\btheta_0}  = \0$  is equivalent to
\[
\mathrm{col}(\H\G_{\t_0\btheta_0}) \subseteq \mathrm{col}(\R)=\mathrm{col}(\H\G_{\t_0\efree_0}), 
\]
which is the same as 
\[
\mathrm{col}(\G_{\t_0\btheta_0}) \subseteq \mathrm{col}(\G_{\t_0\efree_0}),
\]
since $\H$ is invertible. 
This completes the proof because $\G_{\t_0\btheta_0} = \big( \G_{\t_0\shared_0} \ \, \0 \big)$ by the fact that $\G_{\t_0\pfree_0} = \0$.

\subsection{Moment Condition \eqref{eq:adv} or \eqref{eq:adv+}}

Note that \eqref{eq:adv} is a special case of \eqref{eq:adv+}. 
We need to show Assumption 3 is needed for \eqref{eq:adv+}, which is the same as
\begin{equation}\label{r7} 
E\{p_k(\X\mid \bphi) h(\Z) f_1(\X)/f_0(\X)\mid S=0\} =
E\{q_k(\Z) h(\Z) f_1(\X)/f_0(\X)\mid S=0\}, \end{equation}
where 
\[ q_k(\z) = E \{ q_k (\X ) \mid \z , S=0 \} = \int p_k(\x \mid \bphi ) f_0( \u \mid \z )d \u , \]
 $f_0( \u \mid \z )$ is the density of $\U$ conditioned on $\Z$ and $S=0$, and $\U$ is the component of $\X$ not in $\Z$. 
 Let $f_1( \u \mid \z )$ be the density of $\U$ conditioned on $\Z$ and $S=1$, 
  $f_1( \z )$ be the density of $\Z$ conditioned on $S=1$, and 
 $f_0( \z )$ be the density of $\Z$ conditioned on $S=0$. 
The right side of \eqref{r7}  is 
\begin{align*}
    \int q_k( \z ) h(\z ) \frac{f_1(\x )}{f_0(\x )}  f_0(\x ) d \x & = 
    \int q_k(\z ) h(\z ) f_1(\x ) d \x \\
& = \int q_k( \z ) h(\z )  f_1 (\z )d \z \\
& = \int \!\!\! \int p_k( \x \mid \bphi ) f_0( \u \mid \z ) d \u \, h(\z ) f_1(\z )d \z \\
& = \int  \!\!\! \int p_k( \x \mid \bphi ) f_1( \u \mid \z ) d \u \, h(\z ) f_1(\z )d \z \\
& = \int p_k( \x \mid \bphi ) \, h(\z ) f_1(\x ) d \x \\
& =\int p_k( \x \mid \bphi ) \, h(\z ) \frac{f_1(\x )}{f_0(\x )}  f_0(\x ) d \x, 
\end{align*} 
the left side of \eqref{r7}, where the fourth equality follows from 
$ f_0( \u \mid \z )= f_1( \u \mid \z )$ under  Assumption 3. This proof also indicates that if Assumption 3 does not hold, then $ f_0( \u \mid \z ) \neq f_1( \u \mid \z )$ and, thus, the left and right sides of \eqref{r7} are not equal in general. 

\vskip 0.2in

\let\v\vaccent
\bibliography{ref}

\end{document}